\documentclass[conference]{IEEEtran}
\IEEEoverridecommandlockouts

\usepackage{cite}
\usepackage{amsmath,amssymb,amsfonts}
\usepackage{algorithm}
\usepackage{algorithmic}
\usepackage{graphicx}
\usepackage{textcomp}
\usepackage{xcolor}
\usepackage{caption}
\usepackage{comment}
\usepackage{subcaption}
\usepackage{tikz}
\usetikzlibrary{calc}
\usepackage{amsmath,amssymb}
\newcommand{\RR}{\mathbb{R}} % For Real
\usetikzlibrary{arrows.meta}
\usepackage[acronyms,nonumberlist,nopostdot,nomain,nogroupskip,acronymlists={hidden}]{glossaries}
\newglossary[algh]{hidden}{acrh}{acnh}{Hidden Acronyms}
\newacronym{collprob}{CP}{Collision Probability}
\newacronym{pcr}{PCR}{Packet Collision Ratio}
\newacronym{cbr}{CBR}{Channel Bussy Ration}
\newacronym{pir}{PIR}{Packet Inter-Reception}
\newacronym{ud}{UD}{Update Delay}
\newacronym{sr}{SR}{Scheduling Request}
\newacronym{dg}{DG}{Distrubuted Grant}
\newacronym{prr}{PRR}{Packet Reception Ratio}
\newacronym{ma}{MA}{Multi-Agent}
\newacronym{dqn}{DQN}{Deep Q-Network}
\newacronym{madrl}{MADRL}{Multi-Agent Deep Reinforcement Learning}
\newacronym{marl}{MARL}{Multi-Agent Reinforcement Learning}
\newacronym{psbch}{PSBCH}{Physical Sidelink Broadcast Channel}
\newacronym{psfch}{PSFCH}{Physical Sidelink Feedback Channel}
\newacronym{pscch}{PSCCH}{Physical Sidelink Control Channel}
\newacronym{pssch}{PSSCH}{Physical Sidelink Shared Channel}
\newacronym{scs}{SCS}{Subcarrier Spacing}
\newacronym{ndi}{NDI}{New Data Indicator}
\newacronym{rc}{RC}{re-selection counter}
\newacronym{rp}{RP}{Resource Pool}
\newacronym{rri}{RRI}{Resource Reservation Interval}
\newacronym{rsrp}{RSRP}{Reference Signal Received Power}

\newacronym{rnn}{RNN}{Recurrent Neural Networks}
\newacronym{ra}{RA}{Resource Allocation}
\newacronym{dra}{DRA}{Distributed Resource Allocation}
\newacronym{rsvp}{P$_{rsvp}$}{Resource Reservation Period}
\newacronym{sci}{SCI}{Sidelink Control Information}
\newacronym{rl}{RL}{Reinforcement Learning}
\newacronym{ppbp}{PPBP}{Poisson Pareto Burst Process}
\newacronym{pdr}{PDR}{Packet Delivery Ratio}

\newacronym{hu}{HU}{Heavy User}
\newacronym{dnn}{DNN}{Deep-Neural Network}
\newacronym{relu}{ReLU}{Rectified Linear Unit}
\newacronym{ns3}{ns-3}{Network Simulator 3}
\newacronym{sl}{SL}{Sidelink}
\newacronym{rnti}{SL-RNTI}{Sidelink Radio Network Temporary Identifier}
\newacronym{3gpp}{3GPP}{3rd Generation Partnership Project}
\newacronym{4g}{4G}{4th generation}
\newacronym{5g}{5G}{5th generation}
\newacronym{6g}{6G}{6th generation}
\newacronym{5gc}{5GC}{5G Core}
\newacronym{isd}{ISD}{Intersite distance}

\newacronym{adc}{ADC}{Analog to Digital Converter}
\newacronym{aerpaw}{AERPAW}{Aerial Experimentation and Research Platform for Advanced Wireless}
\newacronym{ai}{AI}{Artificial Intelligence}
\newacronym{aimd}{AIMD}{Additive Increase Multiplicative Decrease}
\newacronym{am}{AM}{Acknowledged Mode}
\newacronym{amc}{AMC}{Adaptive Modulation and Coding}
\newacronym{amf}{AMF}{Access and Mobility Management Function}
\newacronym{aops}{AOPS}{Adaptive Order Prediction Scheduling}
\newacronym{api}{API}{Application Programming Interface}
\newacronym{apn}{APN}{Access Point Name}
\newacronym{ap}{AP}{Application Protocol}
\newacronym{aqm}{AQM}{Active Queue Management}
\newacronym{ausf}{AUSF}{Authentication Server Function}
\newacronym{avc}{AVC}{Advanced Video Coding}
\newacronym{awgn}{AGWN}{Additive White Gaussian Noise}
\newacronym{balia}{BALIA}{Balanced Link Adaptation Algorithm}
\newacronym{bbu}{BBU}{Base Band Unit}
\newacronym{bdp}{BDP}{Bandwidth-Delay Product}
\newacronym{ber}{BER}{Bit Error Rate}
\newacronym{bf}{BF}{Beamforming}
\newacronym{bler}{BLER}{Block Error Rate}
\newacronym{brr}{BRR}{Bayesian Ridge Regressor}
\newacronym{bs}{BS}{Base Station}
\newacronym{bsr}{BSR}{Buffer Status Report}
\newacronym{bss}{BSS}{Business Support System}
\newacronym{ca}{CA}{Carrier Aggregation}
\newacronym{caas}{CaaS}{Connectivity-as-a-Service}
\newacronym{cav}{CAV}{Connected and Autonomous Vehicle}
\newacronym{cb}{CB}{Code Block}
\newacronym{cc}{CC}{Congestion Control}
\newacronym{ccid}{CCID}{Congestion Control ID}
\newacronym{cco}{CC}{Carrier Component}
\newacronym{cd}{CD}{Continuous Delivery}
\newacronym{cdd}{CDD}{Cyclic Delay Diversity}
\newacronym{cdf}{CDF}{Cumulative Distribution Function}
\newacronym{cdn}{CDN}{Content Distribution Network}
\newacronym{cli}{CLI}{Command-line Interface}
\newacronym{cn}{CN}{Core Network}
\newacronym{codel}{CoDel}{Controlled Delay Management}
\newacronym{comac}{COMAC}{Converged Multi-Access and Core}
\newacronym{cord}{CORD}{Central Office Re-architected as a Datacenter}
\newacronym{cornet}{CORNET}{COgnitive Radio NETwork}
\newacronym{cosmos}{COSMOS}{Cloud Enhanced Open Software Defined Mobile Wireless Testbed for City-Scale Deployment}
\newacronym{cots}{COTS}{Commercial Off-the-Shelf}
\newacronym{cp}{CP}{Control Plane}
\newacronym{cyp}{CP}{Cyclic Prefix}
\newacronym{up}{UP}{User Plane}
\newacronym{cpu}{CPU}{Central Processing Unit}
\newacronym{cqi}{CQI}{Channel Quality Information}
\newacronym{cr}{CR}{Cognitive Radio}
\newacronym{cran}{CRAN}{Cloud \gls{ran}}
\newacronym{crs}{CRS}{Cell Reference Signal}
\newacronym{csi}{CSI}{Channel State Information}
\newacronym{csirs}{CSI-RS}{Channel State Information - Reference Signal}
\newacronym{cu}{CU}{Central Unit}
\newacronym{d2tcp}{D$^2$TCP}{Deadline-aware Data center TCP}
\newacronym{d3}{D$^3$}{Deadline-Driven Delivery}
\newacronym{dac}{DAC}{Digital to Analog Converter}
\newacronym{dag}{DAG}{Directed Acyclic Graph}
\newacronym{das}{DAS}{Distributed Antenna System}
\newacronym{dash}{DASH}{Dynamic Adaptive Streaming over HTTP}
\newacronym{dc}{DC}{Dual Connectivity}
\newacronym{dccp}{DCCP}{Datagram Congestion Control Protocol}
\newacronym{dce}{DCE}{Direct Code Execution}
\newacronym{dci}{DCI}{Downlink Control Information}
\newacronym{dctcp}{DCTCP}{Data Center TCP}
\newacronym{dl}{DL}{Downlink}
\newacronym{dmr}{DMR}{Deadline Miss Ratio}
\newacronym{dmrs}{DMRS}{DeModulation Reference Signal}
\newacronym{drlcc}{DRL-CC}{Deep Reinforcement Learning Congestion Control}
\newacronym{dsrc}{DSRC}
{Dedicated Short Range Communications}
\newacronym{d2d}{D2D}{device-to-device}
\newacronym{drs}{DRS}{Discovery Reference Signal}
\newacronym{du}{DU}{Distributed Unit}
\newacronym{e2e}{E2E}{end-to-end}
\newacronym{earfcn}{EARFCN}{E-UTRA Absolute Radio Frequency Channel Number}
\newacronym{ecaas}{ECaaS}{Edge-Cloud-as-a-Service}
\newacronym{ecn}{ECN}{Explicit Congestion Notification}
\newacronym{edf}{EDF}{Earliest Deadline First}
\newacronym{embb}{eMBB}{Enhanced Mobile Broadband}
\newacronym{empower}{EMPOWER}{EMpowering transatlantic PlatfOrms for advanced WirEless Research}
\newacronym{enb}{eNB}{evolved Node Base}
\newacronym{endc}{EN-DC}{E-UTRAN-\gls{nr} \gls{dc}}
\newacronym{epc}{EPC}{Evolved Packet Core}
\newacronym{eps}{EPS}{Evolved Packet System}
\newacronym{es}{ES}{Edge Server}
\newacronym{etsi}{ETSI}{European Telecommunications Standards Institute}
\newacronym[firstplural=Estimated Times of Arrival (ETAs)]{eta}{ETA}{Estimated Time of Arrival}
\newacronym{eutran}{E-UTRAN}{Evolved Universal Terrestrial Access Network}
\newacronym{faas}{FaaS}{Function-as-a-Service}
\newacronym{fapi}{FAPI}{Functional Application Platform Interface}
\newacronym{fdd}{FDD}{Frequency Division Duplexing}
\newacronym{fdm}{FDM}{Frequency Division Multiplexing}
\newacronym{fdma}{FDMA}{Frequency Division Multiple Access}
\newacronym{fed4fire}{FED4FIRE+}{Federation 4 Future Internet Research and Experimentation Plus}
\newacronym{fir}{FIR}{Finite Impulse Response}
\newacronym{fit}{FIT}{Future \acrlong{iot}}
\newacronym{fpga}{FPGA}{Field Programmable Gate Array}
\newacronym{fr2}{FR2}{Frequency Range 2}
\newacronym{fr1}{FR1}{Frequency Range 1}
\newacronym{fs}{FS}{Fast Switching}
\newacronym{fscc}{FSCC}{Flow Sharing Congestion Control}
\newacronym{ftp}{FTP}{File Transfer Protocol}
\newacronym{fw}{FW}{Flow Window}
\newacronym{ge}{GE}{Gaussian Elimination}
\newacronym{gnb}{gNB}{Next Generation Node Base}
\newacronym{gop}{GOP}{Group of Pictures}
\newacronym{gpr}{GPR}{Gaussian Process Regressor}
\newacronym{gps}{GPS}{Global Position System}
\newacronym{gpu}{GPU}{Graphics Processing Unit}
\newacronym{gtp}{GTP}{GPRS Tunneling Protocol}
\newacronym{gtpc}{GTP-C}{GPRS Tunnelling Protocol Control Plane}
\newacronym{gtpu}{GTP-U}{GPRS Tunnelling Protocol User Plane}
\newacronym{gtpv2c}{GTPv2-C}{\gls{gtp} v2 - Control}
\newacronym{gw}{GW}{Gateway}
\newacronym{harq}{HARQ}{Hybrid Automatic Repeat reQuest}
\newacronym{hetnet}{HetNet}{Heterogeneous Network}
\newacronym{hh}{HH}{Hard Handover}
\newacronym{hol}{HOL}{Head-of-Line}
\newacronym{hqf}{HQF}{Highest-quality-first}
\newacronym{hss}{HSS}{Home Subscription Server}
\newacronym{http}{HTTP}{HyperText Transfer Protocol}
\newacronym{ia}{IA}{Initial Access}
\newacronym{iab}{IAB}{Integrated Access and Backhaul}
\newacronym{ic}{IC}{Incident Command}
\newacronym{ietf}{IETF}{Internet Engineering Task Force}
\newacronym{imsi}{IMSI}{International Mobile Subscriber Identity}
\newacronym{imt}{IMT}{International Mobile Telecommunication}
\newacronym{iot}{IoT}{Internet of Things}
\newacronym{ip}{IP}{Internet Protocol}
\newacronym{itu}{ITU}{International Telecommunication Union}
\newacronym{kpi}{KPI}{Key Performance Indicator}
\newacronym{kpm}{KPM}{Key Performance Measurement}
\newacronym{kvm}{KVM}{Kernel-based Virtual Machine}
\newacronym{los}{LoS}{Line of Sight}
\newacronym{lsm}{LSM}{Link-to-System Mapping}
\newacronym{lstm}{LSTM}{Long Short Term Memory}
\newacronym{lte}{LTE}{Long Term Evolution}
\newacronym{lxc}{LXC}{Linux Container}
\newacronym{m2m}{M2M}{Machine to Machine}
\newacronym{mac}{MAC}{Medium Access Control}
\newacronym{manet}{MANET}{Mobile Ad Hoc Network}
\newacronym{mano}{MANO}{Management and Orchestration}
\newacronym{mc}{MC}{Multi-Connectivity}
\newacronym{mcc}{MCC}{Mobile Cloud Computing}
\newacronym{mchem}{MCHEM}{Massive Channel Emulator}
\newacronym{mcs}{MCS}{Modulation and Coding Scheme}
\newacronym{mec2}{MEC}{Multi-access Edge Computing}
\newacronym{mec}{MEC}{Mobile Edge Computing}
\newacronym{mfc}{MFC}{Mobile Fog Computing}
\newacronym{mgen}{MGEN}{Multi-Generator}
\newacronym{mi}{MI}{Mutual Information}
\newacronym{mib}{MIB}{Master Information Block}
\newacronym{miesm}{MIESM}{Mutual Information Based Effective SINR}
\newacronym{mimo}{MIMO}{Multiple Input, Multiple Output}
\newacronym{ml}{ML}{Machine Learning}
\newacronym{mlr}{MLR}{Maximum-local-rate}
\newacronym[plural=\gls{mme}s,firstplural=Mobility Management Entities (MMEs)]{mme}{MME}{Mobility Management Entity}
\newacronym{mmtc}{mMTC}{Massive Machine-Type Communications}
\newacronym{mmwave}{mmWave}{millimeter wave}
\newacronym{mpdccp}{MP-DCCP}{Multipath Datagram Congestion Control Protocol}
\newacronym{mptcp}{MPTCP}{Multipath TCP}
\newacronym{mr}{MR}{Maximum Rate}
\newacronym{mrdc}{MR-DC}{Multi \gls{rat} \gls{dc}}
\newacronym{mse}{MSE}{Mean Square Error}
\newacronym{mss}{MSS}{Maximum Segment Size}
\newacronym{mt}{MT}{Mobile Termination}
\newacronym{mtd}{MTD}{Machine-Type Device}
\newacronym{mtu}{MTU}{Maximum Transmission Unit}
\newacronym{mumimo}{MU-MIMO}{Multi-user \gls{mimo}}
\newacronym{mvno}{MVNO}{Mobile Virtual Network Operator}
\newacronym{nalu}{NALU}{Network Abstraction Layer Unit}
\newacronym{nas}{NAS}{Network Attached Storage}
\newacronym{nat}{NAT}{Network Address Translation}
\newacronym{nbiot}{NB-IoT}{Narrow Band IoT}
\newacronym{nfv}{NFV}{Network Function Virtualization}
\newacronym{nfvi}{NFVI}{Network Function Virtualization Infrastructure}
\newacronym{ni}{NI}{Network Interfaces}
\newacronym{nic}{NIC}{Network Interface Card}
\newacronym{now}{NOW}{Non Overlapping Window}
\newacronym{nsm}{NSM}{Network Service Mesh}
\newacronym{nr}{NR}{New Radio}
\newacronym{nrf}{NRF}{Network Repository Function}
\newacronym{nsa}{NSA}{Non Stand Alone}
\newacronym{nse}{NSE}{Network Slicing Engine}
\newacronym{nssf}{NSSF}{Network Slice Selection Function}
\newacronym{o2i}{O2I}{Outdoor to Indoor}
\newacronym{oai}{OAI}{OpenAirInterface}
\newacronym{oaicn}{OAI-CN}{\gls{oai} \acrlong{cn}}
\newacronym{oairan}{OAI-RAN}{\acrlong{oai} \acrlong{ran}}
\newacronym{oam}{OAM}{Operations, Administration and Maintenance}
\newacronym{ofdm}{OFDM}{Orthogonal Frequency Division Multiplexing}
\newacronym{olia}{OLIA}{Opportunistic Linked Increase Algorithm}
\newacronym{omec}{OMEC}{Open Mobile Evolved Core}
\newacronym{onap}{ONAP}{Open Network Automation Platform}
\newacronym{onf}{ONF}{Open Networking Foundation}
\newacronym{onos}{ONOS}{Open Networking Operating System}
\newacronym{oom}{OOM}{\gls{onap} Operations Manager}
\newacronym{opnfv}{OPNFV}{Open Platform for \gls{nfv}}
\newacronym{oran}{O-RAN}{Open RAN}
\newacronym{orbit}{ORBIT}{Open-Access Research Testbed for Next-Generation Wireless Networks}
\newacronym{os}{OS}{Operating System}
\newacronym{oss}{OSS}{Operations Support System}
\newacronym{pa}{PA}{Position-aware}
\newacronym{pase}{PASE}{Prioritization, Arbitration, and Self-adjusting Endpoints}
\newacronym{pawr}{PAWR}{Platforms for Advanced Wireless Research}
\newacronym{pbch}{PBCH}{Physical Broadcast Channel}
\newacronym{pcef}{PCEF}{Policy and Charging Enforcement Function}
\newacronym{pcfich}{PCFICH}{Physical Control Format Indicator Channel}
\newacronym{pcrf}{PCRF}{Policy and Charging Rules Function}
\newacronym{pdcch}{PDCCH}{Physical Downlink Control Channel}
\newacronym{pdcp}{PDCP}{Packet Data Convergence Protocol}
\newacronym{pdsch}{PDSCH}{Physical Downlink Shared Channel}
\newacronym{pdu}{PDU}{Packet Data Unit}
\newacronym{pf}{PF}{Proportional Fair}
\newacronym{pgw}{PGW}{Packet Gateway}
\newacronym{phich}{PHICH}{Physical Hybrid ARQ Indicator Channel}
\newacronym{phy}{PHY}{Physical}
\newacronym{pmch}{PMCH}{Physical Multicast Channel}
\newacronym{pmi}{PMI}{Precoding Matrix Indicators}
\newacronym{powder}{POWDER}{Platform for Open Wireless Data-driven Experimental Research}
\newacronym{ppo}{PPO}{Proximal Policy Optimization}
\newacronym{ppp}{PPP}{Poisson Point Process}
\newacronym{prach}{PRACH}{Physical Random Access Channel}
\newacronym{rb}{RB}{Physical Resource Block}
\newacronym{prb}{PRB}{Physical Resource Block}
\newacronym{psnr}{PSNR}{Peak Signal to Noise Ratio}
\newacronym{pss}{PSS}{Primary Synchronization Signal}
\newacronym{pucch}{PUCCH}{Physical Uplink Control Channel}
\newacronym{pusch}{PUSCH}{Physical Uplink Shared Channel}
\newacronym{qam}{QAM}{Quadrature Amplitude Modulation}
\newacronym{qci}{QCI}{\gls{qos} Class Identifier}
\newacronym{qoe}{QoE}{Quality of Experience}
\newacronym{qos}{QoS}{Quality of Service}
\newacronym{quic}{QUIC}{Quick UDP Internet Connections}
\newacronym{rach}{RACH}{Random Access Channel}
\newacronym{ran}{RAN}{Radio Access Network}
\newacronym{rbg}{RBG}{Resource Block Group}
\newacronym{rcn}{RCN}{Research Coordination Network}
\newacronym{rec}{REC}{Radio Edge Cloud}
\newacronym{red}{RED}{Random Early Detection}
\newacronym{renew}{RENEW}{Reconfigurable Eco-system for Next-generation End-to-end Wireless}
\newacronym{rf}{RF}{Radio Frequency}
\newacronym{rfc}{RFC}{Request for Comments}
\newacronym{rfr}{RFR}{Random Forest Regressor}
\newacronym{ric}{RIC}{RAN Intelligent Controller}
\newacronym{rlc}{RLC}{Radio Link Control}
\newacronym{rlf}{RLF}{Radio Link Failure}
\newacronym{rlnc}{RLNC}{Random Linear Network Coding}
\newacronym{rmr}{RMR}{RIC Message Router}
\newacronym{rmse}{RMSE}{Root Mean Squared Error}
\newacronym{rnis}{RNIS}{Radio Network Information Service}
\newacronym{rr}{RR}{Round Robin}
\newacronym{rrc}{RRC}{Radio Resource Control}
\newacronym{rrm}{RRM}{Radio Resource Management}
\newacronym{rru}{RRU}{Remote Radio Unit}
\newacronym{rs}{RS}{Remote Server}
\newacronym{rsrq}{RSRQ}{Reference Signal Received Quality}
\newacronym{rss}{RSS}{Received Signal Strength}
\newacronym{rssi}{RSSI}{Received Signal Strength Indicator}
\newacronym{rtt}{RTT}{Round Trip Time}
\newacronym{ru}{RU}{Radio Unit}
\newacronym{rus}{RSU}{Road Side Unit}
\newacronym{rw}{RW}{Receive Window}
\newacronym{rx}{RX}{Receiver}
\newacronym{s1ap}{S1AP}{S1 Application Protocol}
\newacronym{sa}{SA}{standalone}
\newacronym{sack}{SACK}{Selective Acknowledgment}
\newacronym{sap}{SAP}{Service Access Point}
\newacronym{sc2}{SC2}{Spectrum Collaboration Challenge}
\newacronym{scef}{SCEF}{Service Capability Exposure Function}
\newacronym{sch}{SCH}{Secondary Cell Handover}
\newacronym{scoot}{SCOOT}{Split Cycle Offset Optimization Technique}
\newacronym{sctp}{SCTP}{Stream Control Transmission Protocol}
\newacronym{sdap}{SDAP}{Service Data Adaptation Protocol}
\newacronym{sdk}{SDK}{Software Development Kit}
\newacronym{sdm}{SDM}{Space Division Multiplexing}
\newacronym{sdma}{SDMA}{Spatial Division Multiple Access}
\newacronym{sdn}{SDN}{Software-defined Networking}
\newacronym{sdr}{SDR}{Software-defined Radio}
\newacronym{seba}{SEBA}{SDN-Enabled Broadband Access}
\newacronym{sgsn}{SGSN}{Serving GPRS Support Node}
\newacronym{sgw}{SGW}{Service Gateway}
\newacronym{si}{SI}{Study Item}
\newacronym{sib}{SIB}{Secondary Information Block}
\newacronym{sinr}{SINR}{Signal to Interference plus Noise Ratio}
\newacronym{sip}{SIP}{Session Initiation Protocol}
\newacronym{siso}{SISO}{Single Input, Single Output}
\newacronym{sla}{SLA}{Service Level Agreement}
\newacronym{sm}{SM}{Service Model}
\newacronym{smo}{SMO}{Service Management and Orchestration}
\newacronym{smsgmsc}{SMS-GMSC}{\gls{sms}-Gateway}
\newacronym{snr}{SNR}{Signal-to-Noise-Ratio}
\newacronym{son}{SON}{Self-Organizing Network}
\newacronym{ngson}{NG SON}{Next Generation Self-Organizing Network}
\newacronym{sptcp}{SPTCP}{Single Path TCP}
\newacronym{srb}{SRB}{Service Radio Bearer}
\newacronym{srn}{SRN}{Standard Radio Node}
\newacronym{srs}{SRS}{Sounding Reference Signal}
\newacronym{ss}{SS}{Synchronization Signal}
\newacronym{sss}{SSS}{Secondary Synchronization Signal}
\newacronym{st}{ST}{Spanning Tree}
\newacronym{svc}{SVC}{Scalable Video Coding}
\newacronym{tb}{TB}{Transport Block}
\newacronym{tcp}{TCP}{Transmission Control Protocol}
\newacronym{tdd}{TDD}{Time Division Duplexing}
\newacronym{tdm}{TDM}{Time Division Multiplexing}
\newacronym{tdma}{TDMA}{Time Division Multiple Access}
\newacronym{tfl}{TfL}{Transport for London}
\newacronym{tfrc}{TFRC}{TCP-Friendly Rate Control}
\newacronym{tft}{TFT}{Traffic Flow Template}
\newacronym{tgen}{TGEN}{Traffic Generator}
\newacronym{tip}{TIP}{Telecom Infra Project}
\newacronym{tm}{TM}{Transparent Mode}
\newacronym{to}{TO}{Telco Operator}
\newacronym{tr}{TR}{Technical Report}
\newacronym{trp}{TRP}{Transmitter Receiver Pair}
\newacronym{ts}{TS}{Technical Specification}
\newacronym{tti}{TTI}{Transmission Time Interval}
\newacronym{ttt}{TTT}{Time-to-Trigger}
\newacronym{tx}{TX}{Transmitter}
\newacronym{uas}{UAS}{Unmanned Aerial System}
\newacronym{uav}{UAV}{Unmanned Aerial Vehicle}
\newacronym{udm}{UDM}{Unified Data Management}
\newacronym{udp}{UDP}{User Datagram Protocol}
\newacronym{udr}{UDR}{Unified Data Repository}
\newacronym{ue}{UE}{User Equipment}
\newacronym{uhd}{UHD}{\gls{usrp} Hardware Driver}
\newacronym{ul}{UL}{Uplink}
\newacronym{um}{UM}{Unacknowledged Mode}
\newacronym{uml}{UML}{Unified Modeling Language}
\newacronym{upa}{UPA}{Uniform Planar Array}
\newacronym{upf}{UPF}{User Plane Function}
\newacronym{urllc}{URLLC}{Ultra Reliable and Low Latency Communications}
\newacronym{usa}{U.S.}{United States}
\newacronym{usim}{USIM}{Universal Subscriber Identity Module}
\newacronym{usrp}{USRP}{Universal Software Radio Peripheral}
\newacronym{utc}{UTC}{Urban Traffic Control}
\newacronym{vim}{VIM}{Virtualization Infrastructure Manager}
\newacronym{vm}{VM}{Virtual Machine}
\newacronym{vnf}{VNF}{Virtual Network Function}
\newacronym{volte}{VoLTE}{Voice over \gls{lte}}
\newacronym{voltha}{VOLTHA}{Virtual OLT HArdware Abstraction}
\newacronym{vr}{VR}{Virtual Reality}
\newacronym{vran}{vRAN}{Virtualized \gls{ran}}
\newacronym{vss}{VSS}{Video Streaming Server}
\newacronym{v2x}{V2X}{Vehicle-to-Everything}
\newacronym{v2i}{V2I}{Vehicle-to-Infrastructure}
\newacronym{v2v}{V2V}{Vehicle-to-Vehicle}
\newacronym{cv2v}{C-V2V}{Cellular-\gls{v2v}}
\newacronym{cv2x}{C-V2X}{Cellular-V2X}
\newacronym{v2n}{V2N}{vehicle-to-network}
\newacronym{wbf}{WBF}{Wired Bias Function}
\newacronym{wf}{WF}{Waterfilling}
\newacronym{wg}{WG}{Working Group}
\newacronym{wlan}{WLAN}{Wireless Local Area Network}
\newacronym{osm}{OSM}{Open Source \gls{nfv} Management and Orchestration}
\newacronym{pnf}{PNF}{Physical Network Function}
\newacronym{drl}{DRL}{Deep Reinforcement Learning}
\newacronym{mtc}{MTC}{Machine-type Communications}
\newacronym{osc}{OSC}{O-RAN Software Community}
\newacronym{mns}{MnS}{Management Services}
\newacronym{ves}{VES}{\gls{vnf} Event Stream}
\newacronym{ei}{EI}{Enrichment Information}
\newacronym{fh}{FH}{Fronthaul}
\newacronym{fft}{FFT}{Fast Fourier Transform}
\newacronym{laa}{LAA}{Licensed-Assisted Access}
\newacronym{plfs}{PLFS}{Physical Layer Frequency Signals}
\newacronym{ptp}{PTP}{Precision Time Protocol}
\newacronym{lidar}{LiDAR}{Light Detection And Ranging}
\newacronym{dem}{DEM}{Digital Elevation Model}
\newacronym{dtm}{DEM}{Digital Terrain Model}
\newacronym{dsm}{DEM}{Digital Surface Models}
\newacronym{ota}{OTA}{Over-The-Air}
\newacronym{ns}{NS}{Network Slicing}
\newacronym{ne}{NE}{Nash Equilibrium}
\newacronym{hf}{HF}{High Frequency}
\newacronym{noma}{NOMA}{Non-Orthogonal Multiple Access}
\newacronym{sre}{SRE}{Smart Radio Environment}
\newacronym{ris}{RIS}{Reconfigurable Intelligent Surface}
\newacronym{inp}{InP}{Infrastructure Provider}
\newacronym{smf}{SMF}{Slicing Magangement Framework}
\newacronym{nsn}{NSN}{Network Slicing Negotiation}
\newacronym{sms}{SMS}{Slicing MAC Scheduler}
\newacronym{brd}{BRD}{Best Response Dynamics}
\newacronym{dssbr}{DSSBR}{Double Step Smoothed Best Response}
\newacronym{poa}{PoA}{Price of Anarchy}
\newacronym{pos}{PoS}{Price of Stability}
\newacronym{milp}{MILP}{Mixed Integer-Linear Program}
\newacronym{pod}{PoD}{Price of DSSBR}
\newacronym{roc}{ROC}{Radio Overload Control}
\newacronym{ciot}{cIoT}{critical Internet of Things}
\newacronym{embbpr}{eMBB Pr.}{enhanced Mobile BroadBand Premium}
\newacronym{sps}{SPS}{Semi-persistent Scheduling}
\newacronym{cg}{CG}{Configured Grant}
\newacronym{embbbs}{eMBB Bs.}{enhanced Mobile BroadBand Basic}
\newacronym{en}{EN}{Edge Node}
\newacronym{ec}{EC}{Edge Computing}
\newacronym{sp}{SP}{Service Provider}
\newacronym{me}{ME}{Market Equilibrium}
\newacronym{so}{SO}{Social Optimum}
\newacronym{wso}{WSO}{Weighted Social Optimum}
\newacronym{ps}{PS}{Proportional Sharing}
\newacronym{eg}{EG}{Eisenberg-Gale program}
\newacronym{pe}{PE}{Pareto Efficiency}
\newacronym{nsw}{NSW}{Nash Social Welfare}
\newacronym{ef}{EF}{Envy-Freeness}
\newacronym{sub6}{sub-6GHz}{Below 6GHz}
\newacronym{ncr}{NCR}{Network-Controlled Repeater}
\newacronym{nlos}{NLoS}{Non-Line of Sight}
\newacronym{src}{SRC}{Smart Radio Connection}
\newacronym{srd}{SRD}{Smart Radio Device}
\newacronym{cs}{CS}{Candidate Site}
\newacronym{tp}{TP}{Test Point}
\newacronym{fov}{FoV}{Field of View}
\newacronym{nrric}{near-RT RIC}{Near Real-time {RAN} Intelligent Controller}
\newacronym{e2ap}{E2AP}{E2 Application Protocol}
\newacronym{e2sm}{E2SM}{E2 Service Model}
\newacronym{nrtric}{non-RT RIC}{Non-Real-Time {RIC}}
\newacronym{itti}{ITTI}{Inter-task Interface}
\newacronym{bap}{BAP}{Backhaul Adaptation Protocol}
\newacronym{iabest}{IABEST}{Integrated Access and Backhaul Experimental large-Scale Tetbed}
\newacronym{teid}{TEID}{Tunnel Endpoint Identifier}
\newacronym{dlsch}{DL-SCH}{Downlink Shared Channel }
\newacronym{ulsch}{UL-SCH}{Uplink Shared Channel }
\newacronym{rsu}{RSU}{Road Side Unit}
\newacronym{its}{ITS}{Intelligent Transportation Systems}
\newacronym{vanet}{VANET}{Vehicular Ad-hoc Network}
\newacronym{dt}{DT}{Digital Twin}
\newacronym{ecc}{ECC}{Edge Computing Cluster}
\newacronym{obu}{OBU}{On Board Unit}
\newacronym{prdr}{PRDR}{Packet Reception Disagreement Ratio}
\newacronym{dr}{DR}{Disagreement Ratio}
\newacronym{ndt}{NDT}{Network Digital Twin}
\newacronym{cam}{CAM}{Cooperative Awareness Message}
\newacronym{cpm}{CPM}{Collective Perception Message}
\newacronym{pdf}{PDF}{Probability Density Function}
\newacronym[
  firstplural=Radio Access Technologies (RATs),
  longplural=Radio Access Technologies
]{rat}{RAT}{Radio Access Technology}
\newacronym{nlosv}{NLoSv}{Non-Line of Sight due to Vehicles}
\newacronym{gnss}{GNSS}{Global Navigation Satellite System}
\newacronym{rtk}{RTK}{Real-Time Kinematic}
\newacronym{sbr}{SBR}{Shooting and Bouncing Rays}
\newacronym{rtdi}{RTDI}{Ray Tracing Density Index}
\newacronym{cir}{CIR}{Channel Impulse Response}
\newacronym{lan}{LAN}{Local Area Network}
\newacronym{di}{DI}{Density Index}

\def\BibTeX{{\rm B\kern-.05em{\sc i\kern-.025em b}\kern-.08em
    T\kern-.1667em\lower.7ex\hbox{E}\kern-.125emX}}
\begin{document}

\title{Predicting Networks Before They Happen: Experimentation on a Real-Time V2X Digital Twin}
\author{Roberto Pegurri\textsuperscript{1},
Shintaro Habu\textsuperscript{2},
Francesco Linsalata\textsuperscript{1},
Kui Wang\textsuperscript{2},
Tao Yu\textsuperscript{2},\\
Eugenio Moro\textsuperscript{1},
Maiya Igarashi\textsuperscript{2},
Antonio Capone\textsuperscript{1}, 
Kei Sakaguchi\textsuperscript{2}
\\
{\textit{\textsuperscript{1}Department of Electronics, Information and Bioengineering, Politecnico di Milano, Italy}}
\\
{\textit{\textsuperscript{2}{Department of Electrical and Electronic Engineering, Institute of Science Tokyo, Japan}}}
\\ 

\small{Email: 
\textsuperscript{1}\{name.surname\}@polimi.it},
\textsuperscript{2}\{habu, kuiw, yutao, igarashi, sakaguchi\}@mobile.ee.titech.ac.jp}

\maketitle

\begin{abstract}
Emerging safety-critical Vehicle-to-Everything (V2X) applications require networks to proactively adapt to rapid environmental changes rather than merely reacting to them. While Network Digital Twins (NDTs) offer a pathway to such predictive capabilities, existing solutions typically struggle to reconcile high-fidelity physical modeling with strict real-time constraints. This paper presents a novel, end-to-end real-time V2X Digital Twin framework that integrates live mobility tracking with deterministic channel simulation. By coupling the Tokyo Mobility Digital Twin---which provides live sensing and trajectory forecasting---with VaN3Twin---a full-stack simulator with ray tracing---we enable the prediction of network performance before physical events occur. We validate this approach through an experimental proof-of-concept deployed in Tokyo, Japan, featuring connected vehicles operating on 60~GHz links. Our results demonstrate the system's ability to predict Received Signal Strength (RSSI) with a maximum average error of 1.01~dB and reliably forecast Line-of-Sight (LoS) transitions within a maximum average end-to-end system latency of 250~ms, depending on the ray tracing level of detail. Furthermore, we quantify the fundamental trade-offs between digital model fidelity, computational latency, and trajectory prediction horizons, proving that high-fidelity and predictive digital twins are feasible in real-world urban environments.
\end{abstract}


\section{Introduction} \label{sec:introduction}
 Connected and automated vehicles are pushing wireless networks to operate increasingly close to their physical limits, both in terms of reliability and reaction time. Safety critical \gls{v2x} applications such as collision avoidance, cooperative perception, and automated intersection management require the network to reason not only about the current state of the environment, but also on how positions and channels will evolve in the immediate future, from tens of milliseconds to a few seconds ahead~\cite{11038674}. 
This is precisely the promise of \glspl{ndt}: a software replica of the physical system, continuously synchronized with live data, that can be queried to evaluate “what–if’’ situations before they affect real users~\cite{9839640}. 
Despite significant progress, most existing \glspl{ndt} operate in an essentially \emph{offline} manner.  
High–fidelity ray tracing, realistic mobility engines, and multi-RAT protocol stacks are routinely used to obtain accurate performance evaluations, but without strict timing constraints, simulation runs may take seconds, minutes, or even hours to complete.
This is acceptable for design-time analysis, but incompatible with an online, predictive \gls{dt}, for which results must be delivered within the tight latency budget~\cite{9429703}, for example as imposed by \gls{v2x} applications. 
If the twin delivers its prediction too late, vehicles will have moved, the channel will have changed, and any recommended action may already be obsolete or even unsafe.
Two recent platforms partially address this gap.  
The \emph{Tokyo Mobility \gls{dt}}---from here on referred to as Mobility DT---developed by the Institute of Science Tokyo, Japan~\cite{10443037} provides a detailed 3D reconstruction of the Ookayama Campus area in Tokyo, Japan. It supports both real-time and replayed trajectories enabling realistic urban-scale mobility scenarios.
Conversely, \emph{VaN3Twin}~\cite{pegurri2025van3twinmultitechnologyv2xdigital} extends the ms-van3t/ns-3 framework~\cite{ms-van3t-journal-2024} with the in-the-loop integration of Sionna RT~\cite{sionna}, thereby enabling deterministic ray-based channel prediction and the implementation of \gls{its} protocols over full-stack realizations of multiple \glspl{rat}. The resulting framework has been validated against on-field measurements, demonstrating close agreement between predicted and observed channel behavior~\cite{pegurri2025van3twinmultitechnologyv2xdigital}.

Each platform solves a different part of the problem: the Mobility DT offers realism in geometry and mobility, while VaN3Twin offers realism across the complete communication stack.  
Neither, however, provides an end-to-end real-time pipeline capable of ingesting live mobility, predicting near-future trajectories, computing near-future channels, and take action within a strict deadline.
%In this paper we take a step toward this vision by \emph{integrating VaN3Twin into the Tokyo Mobility DT ecosystem} and by explicitly designing the interaction between the two around real-time constraints. The Tokyo Mobility DT supplies accurate, time-stamped positions and 3D geometry; VaN3Twin uses this information to compute full-stack, ray-traced V2X performance. We organize the interaction as a closed-loop predictive workflow: the Tokyo Mobility DT continuously receives fresh measurements from the physical testbed, predicts short-term mobility, and triggers VaN3Twin to compute \emph{future} channel states within a strict computational budget. If the prediction arrives too late, it is discarded, as it no longer reflects the physical future it was meant to represent.

\subsection{State of the art}

General-purpose network simulators such as ns-3 and OMNeT++ form the foundation of most V2X simulation frameworks. Among them, ms-van3t is one of the most comprehensive: it supports IEEE~802.11p, LTE-V2X, NR-V2X, and the full ETSI ITS-G5 facilities layer, and integrates with SUMO and CARLA for mobility simulation~\cite{ms-van3t-journal-2024}.  
%While widely adopted for V2X research and hardware-in-the-loop experiments, its propagation models remain primarily stochastic or semi-stochastic and are therefore not tailored to site-specific DT operation. 
Several recent works interpret such simulators as components of broader vehicular \glspl{ndt}. 
Examples include blockage-aware mmWave \glspl{dt} for multi-hop topologies~\cite{roongpraiwan2025digital} and multi-modal \glspl{dt} combining mobility, sensing, and coarse communication abstractions~\cite{CazzellaV2XDT}. These platforms move beyond offline simulation but still rely on simplified PHY abstractions to maintain responsiveness. 
Traditional wireless propagation models in ns-3 and OMNeT++ are stochastic (e.g., 3GPP~TR~36.885/38.901) or based on analytical families such as WINNER, able to capture large-scale trends but not the site-specific multipath structure.  
This limitation becomes critical in DT settings, where small changes in geometry or mobility strongly influence short-term channel evolution. To overcome this, several works integrate external ray-based channel simulators into network simulators.  
MATLAB-based approaches~\cite{matlab-rt-colosseum} emphasize PHY accuracy and hardware emulation, but lack full V2X stacks and exhibit poor scalability.  
Other works import offline ray-traced maps into ns-3~\cite{gaugel2012accurate}, enabling bit-level simulation but preventing synchronous updates.  
OPAL--OMNeT++ integration~\cite{RUZNIETO2023100964} achieves realistic geometry but targets non-V2X technologies.
VaN3Twin, however, advances this direction by coupling ms-van3t with Sionna RT into a comprehensive V2X \gls{ndt} framework, enabling high-fidelity, in-the-loop ray-traced propagation and realistic modeling of multi-RAT coexistence over shared spectrum portions. However, despite its channel realism, VaN3Twin has primarily been used offline or in emulation mode, without enforcing a strict real-time prediction deadline.
GPU-accelerated ray tracing software such as Sionna RT now enable repeated evaluations with competitive runtime~\cite{zhu2024toward}.
This has motivated new in-the-loop architectures, including ns3sionna~\cite{ns3-sionna-falko} for Wi-Fi and our prior ns-3 + Sionna RT integration~\cite{Pegu2505:Toward} for multi-RAT 6G studies.
DT-CoVeSS~\cite{Twardokus2505:DT-CoVeSS} similarly uses Sionna RT for high-fidelity analysis of CV2X security.  
These works demonstrate that in-the-loop ray tracing integration is feasible, but none targets city-scale mobility, full V2X communication stacks, and strict prediction deadlines simultaneously. Vehicular \glspl{dt} outside ns-3 focus increasingly on near real-time operation.
RAVEN~\cite{RAVEN-paper} combines live data and contextual information to predict mmWave conditions and support interference management. Vehicle-to-Cloud ADAS~\cite{V2C-ADAS} demonstrates cloud-assisted actuation using DT logic.  
Large-scale mobility \glspl{dt} such as TuST~\cite{TuST-paper-journal} and indoor crowd models~\cite{9861008} show that realistic spatio-temporal digital replicas can be built from real data.
However, these systems often rely on simplified communication models or separate mobility and communication loops.  
%What is missing is a DT that combines: (i) realistic, city-scale mobility (e.g., Tokyo Mobility DT); (ii) a full-stack, ray-traced V2X simulator (VaN3Twin/ms-van3t); and (iii) an explicit timing budget ensuring that predictions are delivered before the corresponding physical time occurs.
What is missing is a DT that combines:  
(i) realistic, city-scale mobility;  
(ii) a full-stack, ray-traced V2X simulator; and  
(iii) an explicit timing budget ensuring that predictions are delivered before the event takes place in the physical world.

%In this work, we address this gap by integrating VaN3Twin within the Tokyo Mobility \gls{dt}, using Sionna RT as a shared ray tracing core, and orchestrating data exchange and scheduling to enable short-term \emph{future} channel prediction at latencies compatible with real-time V2X operation. This integration exposes the fundamental bottlenecks of real-time \glspl{ndt}: (i) the computational cost of the physical-layer model, especially ray-traced propagation; (ii) the overhead of synchronizing mobility and geometry between Tokyo Mobility DT and VaN3Twin; and (iii) the need to schedule operations so that the predicted state becomes available \emph{before} the corresponding physical time is reached. Our goal is not to propose new trajectory or channel prediction algorithms---as these are well studied---but rather to quantify how far a real system can be pushed toward real-time predictive operation when combining a live mobility testbed with a high-fidelity V2X DT. 

\textbf{Paper Contributions} To address the gap identified above, we integrate VaN3Twin within the Tokyo Mobility \gls{dt} ecosystem using Sionna RT as a shared ray tracing core, orchestrating data exchange and scheduling to enable short-term \emph{future} channel prediction at latencies compatible with real-time V2X operation. Any predictions that arrive too late are discarded to maintain consistency with the evolving physical system. This integration exposes the key challenges of real-time NDT operation, including the computational cost of ray-traced propagation, the synchronization of mobility and geometry between platforms, and the scheduling required to deliver actionable predictions before the corresponding physical time occurs. Rather than proposing new trajectory or channel prediction algorithms, our goal is to quantify how far a real system can be pushed toward real-time predictive operation when combining a live mobility testbed with a high-fidelity V2X DT.

The main contributions of this work are as follows:
\begin{itemize}
    \item We design and implement an end-to-end integration of VaN3Twin and the Tokyo Mobility DT into a latency-aware and predictive \gls{ndt};
    \item We introduce a timing-constrained workflow for future channel state prediction, where mobility updates, 3D scene synchronization, ray tracing, and network simulation are orchestrated to meet a strict latency budget;
    \item We demonstrate the integrated system in an urban testbed, quantifying the trade-offs between trajectory prediction error, channel prediction accuracy, and system latency. %and identifying the operational regime where the Digital Twin remains useful for time-critical V2X applications.
\end{itemize}

\textbf{Paper Organization} The remainder of the paper is structured as follows.
Section~\ref{sec:integration} presents the integration architecture and the real-time workflow.  
Section~\ref{sec:e2e_analysis} formalizes how model complexity affects computation time and end-to-end system latency.  
Section~\ref{sec:poc} reports experimental results from the physical deployment in the Ookayama Campus of the Institute of Science Tokyo, Japan.  
Section~\ref{sec:conclusion} concludes the paper.

\section{Integration architecture} \label{sec:integration}
\begin{figure} [b]
    \centering
    \includegraphics[width=1\linewidth]{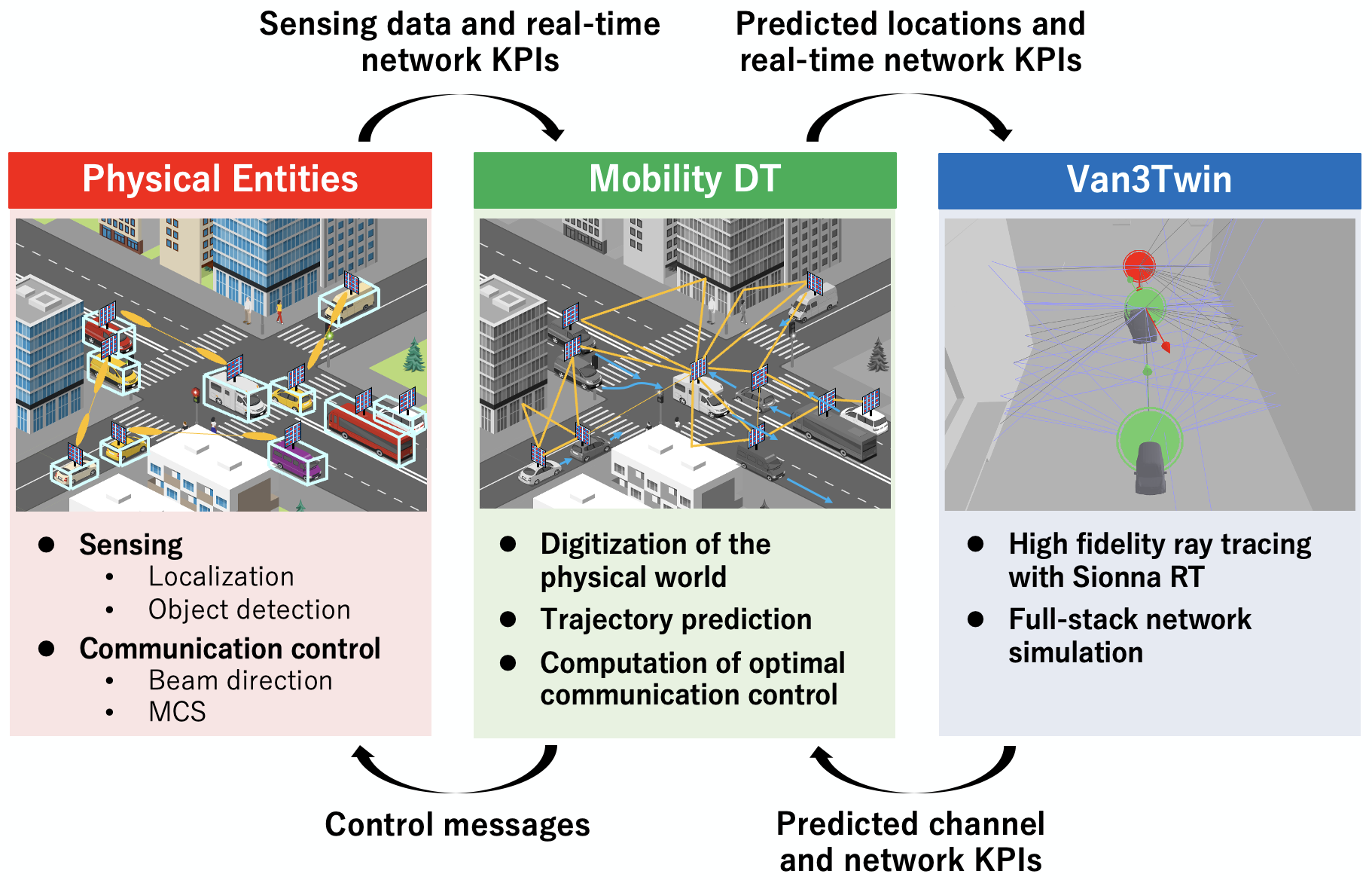}
    \caption{Schematic architecture of the framework.}
    \label{fig:architecture}
\end{figure}

This section presents the architecture of the proposed framework and details the interactions among its main components. 
%Particular emphasis is placed on how measurement data is acquired, processed, and exchanged, as well as on the mechanisms that enable low-latency communication suitable for real-time \gls{ndt} operation. 
As shown in Figure~\ref{fig:architecture}, the framework integrates three core components:
(i) The \emph{physical entities}, i.e., the vehicles equipped with sensing and communication capabilities;
(ii) The \emph{Mobility \gls{dt}}, which maintains a live digital replica of the surrounding traffic and predicts short-term future mobility states;
(iii) \emph{VaN3Twin}, the \gls{ndt} engine responsible for full-stack replication and prediction of V2X communication processes.
%Each component exposes a time-stamped interface to the others, with clearly separated responsibilities. Physical Entities stream raw measurements to the Tokyo Mobility \gls{dt} in the form of tuples $(\mathbf{p}_t, \mathbf{v}_t, \psi_t)$, where $\mathbf{p}_t$, $\mathbf{v}_t$, and $\psi_t$ represent position, velocity, and heading at time $t$, respectively; when available, basic link-layer indicators for time $t$ (e.g., RSSI or throughput) are also attached. Based on these inputs, the Tokyo Mobility \gls{dt} generates short-horizon mobility predictions and exports to VaN3Twin batches of predicted vehicle poses at a target future time $t+h$. VaN3Twin consumes this information to perform the simulation and returns to the Tokyo Mobility \gls{dt} per-link communication KPIs explicitly associated with the predicted time instant $t+h$.
Physical entities provide time-stamped measurements to the Mobility \gls{dt}, which produces short-term mobility predictions for VaN3Twin. VaN3Twin uses these predictions to simulate future communication states and returns per-link KPIs.
In this architecture, the Mobility \gls{dt} serves as the coordinating component of the system. It aggregates real-time mobility and perception data collected by LiDAR-equipped physical entities as well as by LiDAR and camera sensors mounted on nearby \glspl{rsu} or, eventually, network-enabled positioning techniques~\cite{10644093}. Using this information, it continuously maintains and forecasts the mobility state of all dynamic elements in the scenario. Mobility updates are processed at a fixed, pre-defined rate, and predictions are generated for a single, short-term horizon $h$, selected to remain compatible with the end-to-end latency budget of the system. The prediction horizon is kept fixed during operation to ensure deterministic scheduling and bounded computation time. The Mobility \gls{dt} then orchestrates network-level prediction requests to VaN3Twin and subsequently disseminates predicted communication states back to the physical entities---possibly including control commands to be applied to the network configuration.
Communication between system elements is structured according to their functional roles. A specific physical entity provides the interface between physical entities and the Mobility \gls{dt}, leveraging a 2.4~GHz IEEE~802.11n Wi-Fi \gls{cp} for low-latency over-the-air control operations. Meanwhile, the Mobility \gls{dt} and VaN3Twin communicate over a wired Ethernet \gls{lan}, reflecting their co-location at the network edge to minimize latency. All inter-component exchanges employ lightweight UDP messages with JSON-formatted payloads, enabling efficient and low-overhead transmission of state and control information. Message delivery is best-effort: occasional packet losses are tolerated and naturally compensated for by subsequent updates, avoiding retransmission delays. All components rely on a shared time base, with timestamps carried in each message and clocks synchronized using standard network time synchronization mechanisms, ensuring consistent temporal alignment across mobility and communication predictions. The detailed data exchange loop is described in the following subsection.

\subsection{End-to-end Workflow}
\begin{figure}
    \centering
    \includegraphics[width=1\linewidth]{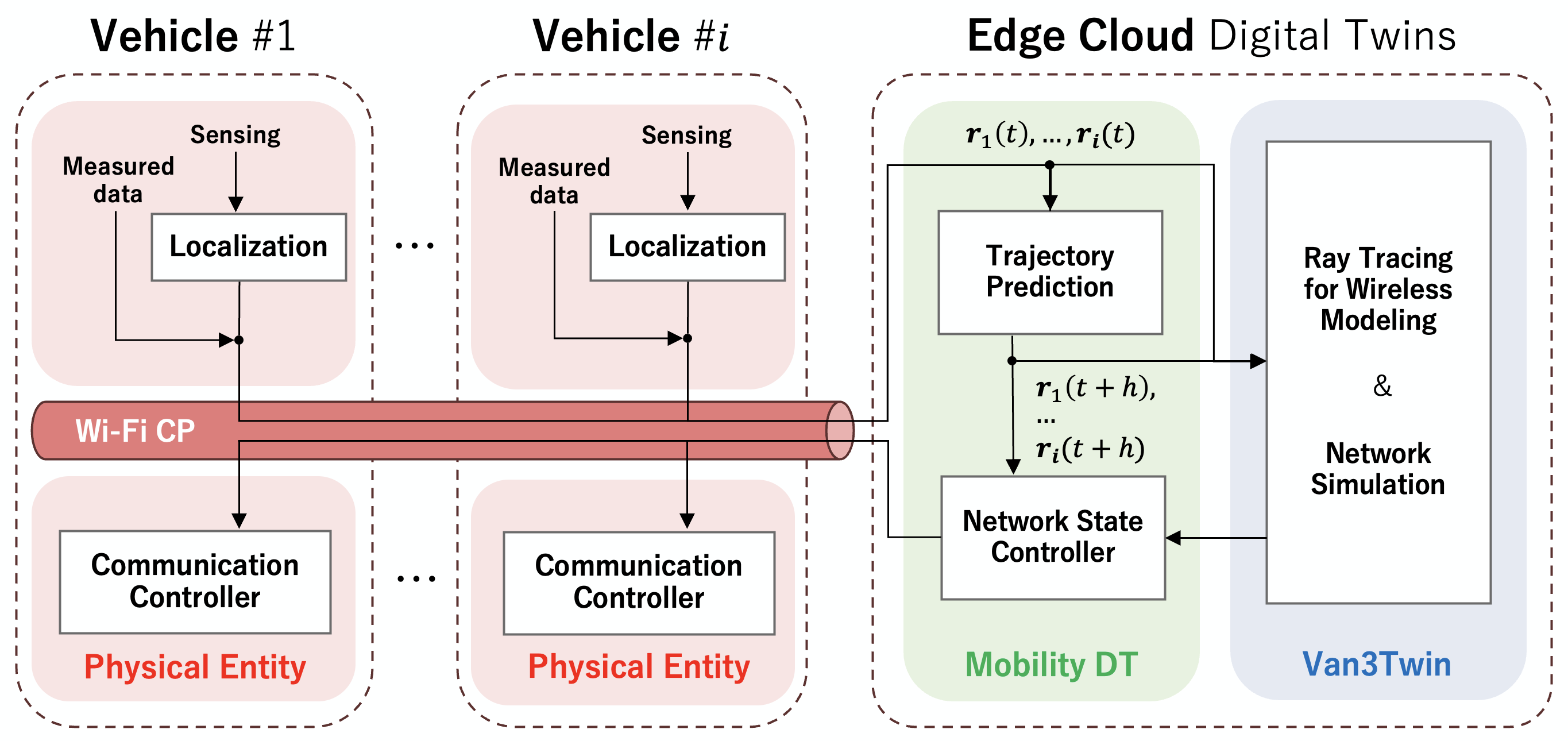}
    \caption{End-to-end communication workflow.}
    \label{fig:workflow}
\end{figure}

The end-to-end workflow illustrated in Figure~\ref{fig:workflow} specifies all the interactions introduced above and clarifies how mobility prediction and communication prediction are coupled in practice. Physical entities periodically transmit their sensed location and measured communication data to the Mobility \gls{dt}---located in the edge cloud---with an implementation-dependent reporting period $\Delta t_{\mathrm{PE}}$. These updates are time-stamped and provide the input required to keep the digital replica synchronized with the physical system and to continuously adjust the internal prediction models.
Based on the aggregated and synchronized mobility state, the Mobility \gls{dt} computes short-term trajectory forecasts for all relevant dynamic objects. Network prediction requests to VaN3Twin are then generated according to a configurable policy, which may be periodic or event-driven, and always target a single future time instant $t+h$. The prediction horizon $h$ is chosen so as to remain compatible with the end-to-end latency budget; multi-horizon requests are also possible in principle but are outside the scope of this work.
Upon receiving a request, VaN3Twin generates a future communication scenario corresponding to the target time $t+h$. Predicted vehicle poses are used to construct a consistent 3D scene, which serves as input to the deterministic ray-based channel model formalized in Section~\ref{sec:e2e_analysis}. The resulting channel realizations drive full-stack V2X simulations, possibly yielding protocol- and packet-level performance indicators. In parallel, the recently observed communication measurements are used to compensate for systematic prediction biases, such as offsets in the computed received power.
The predicted communication state is then returned to the Mobility \gls{dt} and made available for network-level decision making. When required, the DT translates these predictions into control actions which are conveyed back to the physical entities over the dedicated Wi-Fi \gls{cp}, thereby closing the loop between the digital and physical domains.
To reason about the limits of this approach, the next section provides a formal characterization of the proposed integration, detailing the underlying trajectory prediction model, the channel estimation process, and the resulting trade-offs between model complexity, end-to-end latency, and the maximum achievable prediction horizon.

\section{End-to-End Analysis} \label{sec:e2e_analysis}
%This section provides a formal characterization of the integration presented in Section~\ref{sec:integration}. We first describe the vehicle trajectory prediction model, which is used to infer short-term future positions of dynamic entities. We then formalize the ray-based channel prediction process and highlight how uncertainties in trajectory prediction propagate into channel estimation. Finally, we introduce a compact notion of model complexity and show how it affects the end-to-end latency, thereby constraining the maximum feasible prediction horizon.

This section provides a formal characterization of the integration presented in Section~\ref{sec:integration}. We describe the trajectory prediction model, formalize the ray-based channel prediction process, and introduce a concise characterization of model complexity that links physical-model fidelity to end-to-end latency and feasible prediction horizons.

\subsection{Mobility and Channel Prediction} \label{sec:mobility-channel-prediction}
Each vehicle $i$ in the scenario is represented by the kinematic state
$\mathbf{x}^i_t = (\mathbf{p}^i_t,\, \mathbf{v}^i_t,\, \psi^i_t)$.
%where $\mathbf{p}^i_t = (x^i_t, y^i_t)$ denote planar coordinates, $\mathbf{v}^i_t$ the speed, and $\psi^i_t$ the heading angle at time~$t$. 
The state evolution assumes a constant-velocity motion model, with $\mathbf{F}^i_t$ defined accordingly based on the sampling interval. $\mathbf{z}^i_t$ represents the state measurement vector, which the \gls{dt} fuses into a unified state estimate.
%The measurement vector comprises position, velocity, and heading estimates, while the filter state is limited to position only; velocity and heading are treated as auxiliary observations.
The temporal evolution of each vehicle follows a linear state-space model defined by the state-transition matrix $\mathbf{F}^i_t$, the control-input matrix $\mathbf{B}^i_t$, and the measurement model matrix $\mathbf{M}^i_t$, with process noise covariance $\mathbf{Q}^i_t$ and measurement noise covariance $\mathbf{R}^i_t$. The process noise covariance $\mathbf{Q}^i_t$ accounts for deviations from ideal constant-velocity motion and is treated as a design parameter, whereas the measurement noise covariance $\mathbf{R}^i_t$ is sensor-dependent and reflects the heterogeneous accuracy of the contributing sensing modalities.
%Given the filtered estimate $\hat{\mathbf{x}}^{\,i}_{t-1|t-1}$ and covariance $\mathbf{P}^i_{t-1|t-1}$, the \gls{dt} executes a prediction–update cycle---Algorithm~\ref{alg:traj-i}---and then propagates the estimate $h$ steps forward to obtain the predicted pose $\hat{\mathbf{x}}^{\,i}_{t+h|t}$, which is used to instantiate the forecasted traffic geometry.
Given the filtered estimate at time $t$, the \gls{dt} propagates the state $h$ steps forward to obtain the predicted pose $\hat{\mathbf{x}}^{\,i}_{t+h|t}$, which defines the forecasted traffic geometry used for channel prediction (Algorithm~\ref{alg:traj-i}).
\begin{algorithm}[t]
\caption{Trajectory Prediction for Vehicle $i$}
\label{alg:traj-i}
\begin{algorithmic}[1]
    \REQUIRE $\hat{\mathbf{x}}^{\,i}_{t-1|t-1}$, $\mathbf{P}^i_{t-1|t-1}$, $\mathbf{z}^i_t$, $h$
    %\vspace{2pt}
    \STATE Prediction:
        %\vspace{0.5pt}
        \STATE $~~~\hat{\mathbf{x}}^{\,i}_{t|t-1} = \mathbf{F}^i_t \hat{\mathbf{x}}^{\,i}_{t-1|t-1} + \mathbf{B}^i_t \mathbf{u}^i_t $
        %\vspace{1.5pt}
        \STATE $~~~\mathbf{P}^i_{t|t-1} = \mathbf{F}^i_t \mathbf{P}^i_{t-1|t-1} (\mathbf{F}^i_t)^\top + \mathbf{Q}^i_t $
        %\vspace{2pt}
    \STATE Update:
        %\vspace{0.5pt}
        \STATE $~~~\mathbf{K}^i_t = \mathbf{P}^i_{t|t-1} (\mathbf{M}^i_t)^\top \big( \mathbf{M}^i_t \mathbf{P}^i_{t|t-1} (\mathbf{M}^i_t)^\top + \mathbf{R}^i_t \big)^{-1}$
        %\vspace{1.5pt}
        \STATE $~~~\hat{\mathbf{x}}^{\,i}_{t|t} = \hat{\mathbf{x}}^{\,i}_{t|t-1} + \mathbf{K}^i_t \big( \mathbf{z}^i_t - \mathbf{M}^i_t \hat{\mathbf{x}}^{\,i}_{t|t-1} \big)$
        %\vspace{1.5pt}
        \STATE $~~~\mathbf{P}^i_{t|t} = (\mathbf{I} - \mathbf{K}^i_t \mathbf{M}^i_t)\, \mathbf{P}^i_{t|t-1}$
        %\vspace{2pt}
    \STATE Horizon propagation:
        %\vspace{0.5pt}
        \STATE $~~~\hat{\mathbf{x}}^{\,i}_{t+h|t} = (\mathbf{F}^i_t)^h \hat{\mathbf{x}}^{\,i}_{t|t} $
        %\vspace{1.5pt}
    \STATE \textbf{Return:} predicted pose $\hat{\mathbf{x}}^{\,i}_{t+h|t}$
\end{algorithmic}
\end{algorithm}
Once the future scene configuration is assembled, the \gls{dt} evaluates the wireless channel through a deterministic ray-based propagation model.
For each transmitter–receiver pair $(i,j)$, the ray tracing engine emits $N_{\mathrm{rays}}$ rays from the antenna of vehicle~$i$, with departure directions covering the unit sphere.
Each ray propagates through the environment and may undergo up to $N_{\mathrm{int}}$ interactions with surrounding surfaces.
A ray reaching the antenna of vehicle $j$ is part of a valid propagation path that is added to the set $\mathcal{P}^{i,j}_{\mathrm{valid}}$. Each path $p\in\mathcal{P}^{i,j}_{\mathrm{valid}}$ is characterized by its length $d_p^{i,j}$, delay $\tau_p^{i,j}=d_p^{i,j}/c$,
angles of departure $(\theta^{(p)}_{\mathrm{T}},\varphi^{(p)}_{\mathrm{T}})$ and arrival $(\theta^{(p)}_{\mathrm{R}},\varphi^{(p)}_{\mathrm{R}})$,
and by an ordered sequence of $N_p$ interactions.
Its effect on the received field is described by the $2\times 2$ \emph{propagation matrix}:
\begin{equation}
    \mathbf{A}_{p}^{i,j}
    =
    \left(\frac{\lambda}{4\pi d_p^{i,j}}\right)
    e^{-j\frac{2\pi}{\lambda} d_p^{i,j}}
    \prod_{\ell=1}^{N_p}
    \mathbf{M}_{p}^{(\ell)},
    \label{eq:prop-matrix}
\end{equation}
where each $\mathbf{M}_{p}^{(\ell)}$ models the polarization-dependent transformation induced by the $\ell$-th interaction.
The transmit and receive antennas are represented through their complex pattern vectors
$\mathbf{c}_{\mathrm{T}}(\theta,\varphi)$ and $\mathbf{c}_{\mathrm{R}}(\theta,\varphi)$.
The scalar baseband contribution of path $p$ to the channel between vehicles $i$ and $j$ is then 
\begin{equation}
    g_{p}^{i,j}
    =
    \mathbf{c}^{\mathsf{H}}_{\mathrm{R}}
        \!\left(\theta^{(p)}_{\mathrm{R}},\varphi^{(p)}_{\mathrm{R}}\right)
    \mathbf{A}_{p}^{i,j}
    \mathbf{c}_{\mathrm{T}}
        \!\left(\theta^{(p)}_{\mathrm{T}},\varphi^{(p)}_{\mathrm{T}}\right).
    \label{eq:path-gain}
\end{equation}

The predicted multipath channel at time $t+h$ is therefore
\begin{equation}
    h^{i,j}_{t+h|t}(\tau)
    =
    \sum_{p\in\mathcal{P}^{i,j}_{\mathrm{valid}}}
    g_{p}^{i,j}\,
    \delta\!\left(\tau - \tau_p^{i,j}\right).
    \label{eq:channel}
\end{equation}

Collecting all pairwise impulse responses yields the predicted channel matrix
\(
    \widehat{\mathbf{H}}_{t+h|t}
    =
    \big( h^{i,j}_{t+h|t} \big)_{i,j\in\mathcal{V}}.
\)

\subsection{Computational Complexity and Execution Time} \label{sec:complexity-and-time}

\begin{table}
\centering
\caption{Ray tracing parameters for each \gls{di}.}
\begin{tabular}{r c c c c c}
\noalign{\smallskip}
%\textbf{} & \textbf{} & \textbf{} & \textbf{DI} & \textbf{} & \textbf{}\\
\textbf{RT Parameter} & \textbf{DI~1} & \textbf{DI~2} & \textbf{DI~3} & \textbf{DI~4} & \textbf{DI~5}\\
\noalign{\smallskip}
\hline
\noalign{\smallskip }
    %\smallskip
    $\#$~Max. Interactions     & 3            & 3            & 5            & 8            & 8            \\
    %\smallskip
    $\#$~Rays per source       & $10^3$       & $10^3$       & $10^6$       & $10^{10}$    & $10^{10}$    \\
    %\smallskip
    Direct LoS Path            & $\checkmark$ & $\checkmark$ & $\checkmark$ & $\checkmark$ & $\checkmark$ \\
    %\smallskip
    Specular reflection        & $\times$     & $\checkmark$ & $\checkmark$ & $\checkmark$ & $\checkmark$ \\
    %\smallskip
    Diffuse reflection         & $\times$     & $\times$     & $\checkmark$ & $\checkmark$ & $\checkmark$ \\
    %\smallskip
    Refraction                 & $\times$     & $\times$     & $\times$     & $\checkmark$ & $\checkmark$ \\
    %\smallskip
    Diffraction                & $\times$     & $\times$     & $\times$     & $\times$     & $\checkmark$ \\
    \noalign{\smallskip}
    \hline
\end{tabular}
\label{tab:di}
\end{table}
Let $\tau_{e2e}$ be the total time elapsed between the acquisition of a measurement by a vehicle and the completion of the corresponding \gls{dt} update. Formally:
%Following the flow illustrated in Fig.~\ref{fig:e2e-latency-diagram}, it can be expressed as
\begin{align}
    \tau_{e2e}
    = \tau_{m} + \tau_{w} + \tau_{tp}
      + \tau_{req} + \tau_{rt}
      + \tau_{req} + \tau_{w},
      \label{eq:e2e-latency}
\end{align}
where $\tau_m$ is the measurement acquisition time, $\tau_w$ the communication delay between vehicles and the Mobility DT, $\tau_{tp}$ the Mobility DT update and trajectory prediction time, $\tau_{req}$ the inter-process communication delay, and $\tau_{rt}$ the execution time of the physical-modeling stage. 
Given the computational complexity of the ray tracing operation, $\tau_{rt}$ is the dominant and most variable component. 
%Specifically, it depends on the level of physical detail included in the model, such as the number of propagation phenomena considered, the resolution of the spatial sampling, and the depth of the multipath reconstruction. 
Specifically, it depends on the level of physical detail included in the propagation model.
To describe these configurations in a compact and general manner, we introduce the \emph{Ray Tracing Detail Index}, denoted \gls{di}, a scalar metric used to classify the fidelity level of the physical-modeling stage. 
Each \gls{di} level corresponds to a specific configuration of the physical model, ranging from lightweight approximations to high-fidelity representations.  
Table~\ref{tab:di} reports the five \gls{di} levels considered in this work. 
Lower levels include only the essential components needed for coarse propagation estimates, whereas higher levels progressively incorporate more detailed interaction mechanisms, finer spatial resolution, and higher-order multipath modeling.  
Although in our implementation these fidelity levels map to different configurations of a ray-based propagation model, the abstraction provided by \gls{di} remains general and applicable to any physical-modeling backend that supports multiple complexity tiers. 
As the \gls{di} increases, the execution time of the physical model, denoted $\tau_{rt}$, grows accordingly.  
%Since the \gls{dt} predicts the channel for a future time $t+h$, the prediction horizon must satisfy be greater than the required end-to-end system latency, ensuring that the computed prediction is delivered before the physical system diverges from the predicted state. Higher-complexity configurations thus require larger prediction horizons, but larger horizons inherently amplify the trajectory prediction error, which then propagates into the channel estimate. This leads to a fundamental trade-off between model fidelity, computational cost, and predictive accuracy.

To ensure causality, the prediction horizon $h$ must exceed the end-to-end latency $\tau_{e2e}$. Increasing model fidelity therefore requires larger horizons, which in turn amplify trajectory prediction errors and their impact on channel estimation, revealing a fundamental trade-off between physical-model complexity, latency, and predictive accuracy.
Section~\ref{sec:poc} provides an empirical characterization of these trade-offs based on real-world experiments conducted in Tokyo, Japan, evaluating how different \gls{di} levels impact both computational latency and channel prediction accuracy.

\section{Experimental and Simulated Results} \label{sec:poc}
To demonstrate the practical operation of the proposed system, we consider a proof-of-concept deployment carried out in the Ookayama campus of the Institute of Science Tokyo, Japan including measured end-to-end latencies.
To complement the experimental analysis, we then consider a simulation study performed in the identical geometric scenario in which we systematically vary the \gls{di} and introduce synthetic perturbations into the trajectory prediction module.

\subsection{Scenario introduction}

\begin{figure*}
    \centering
    \begin{subfigure}[t]{0.32\textwidth} 
        \centering
        \includegraphics[width=\linewidth]{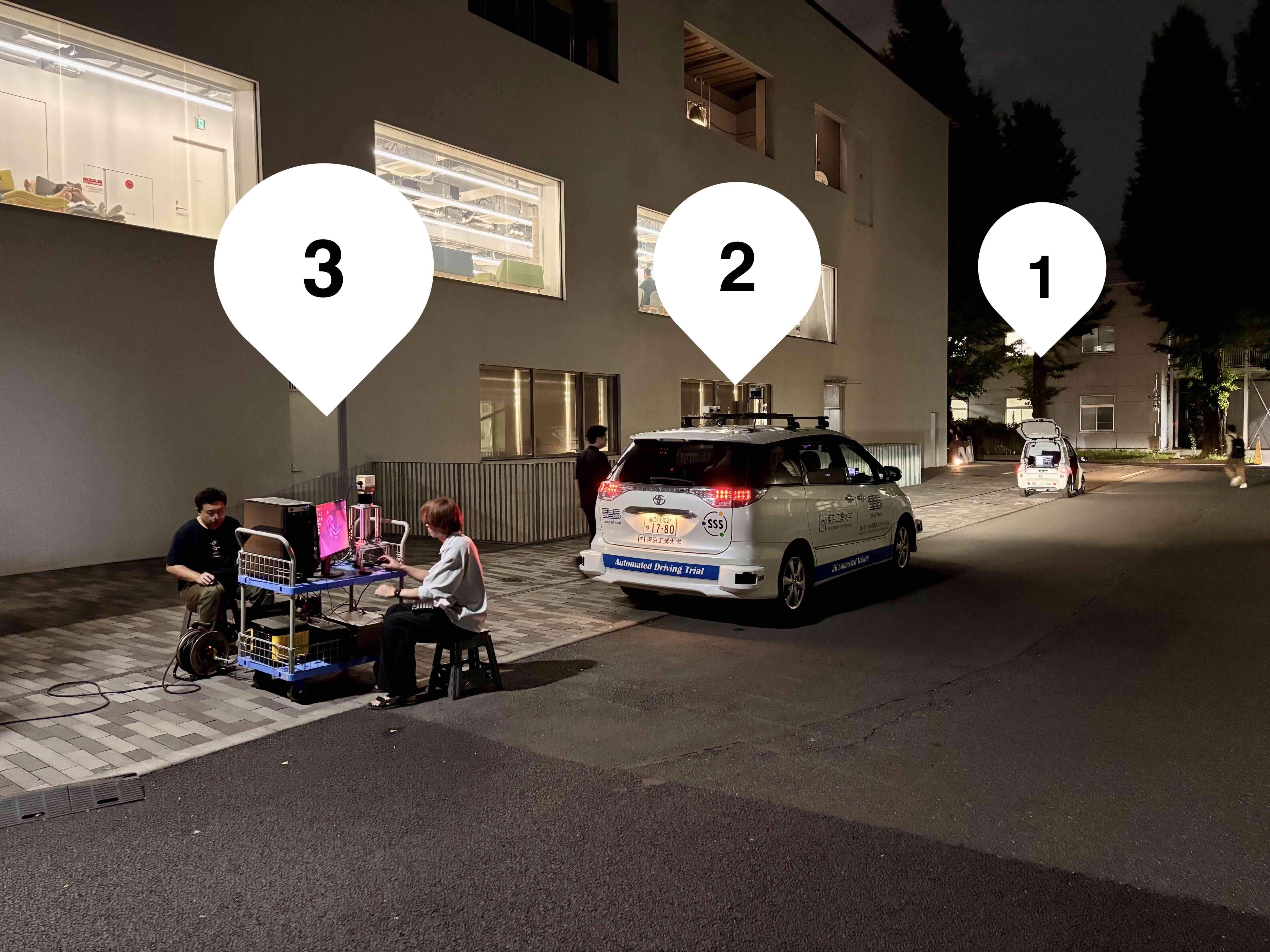} 
        \caption{Real picture of the scenario.}
        \vspace{5pt}
        \label{fig:real-poc}
    \end{subfigure}
    \hfill
    \begin{subfigure}[t]{0.32\textwidth} 
        \centering
        \includegraphics[width=\linewidth]{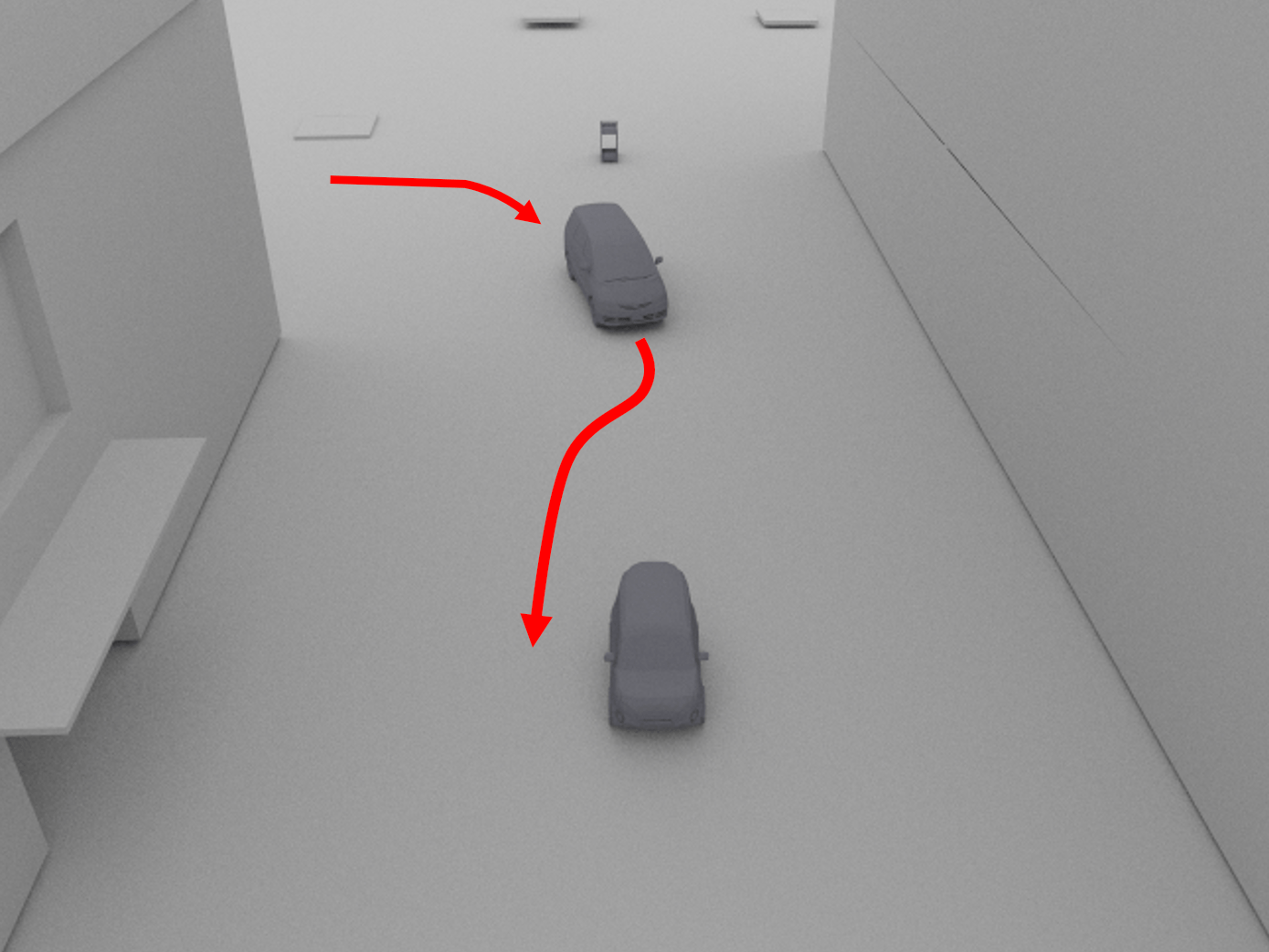} 
        \caption{VaN3Twin scenario with trajectory.}
        \vspace{5pt}
        \label{fig:sionna-poc}
    \end{subfigure}
    \hfill
    \begin{subfigure}[t]{0.32\textwidth} 
        \centering
        \includegraphics[width=\linewidth]{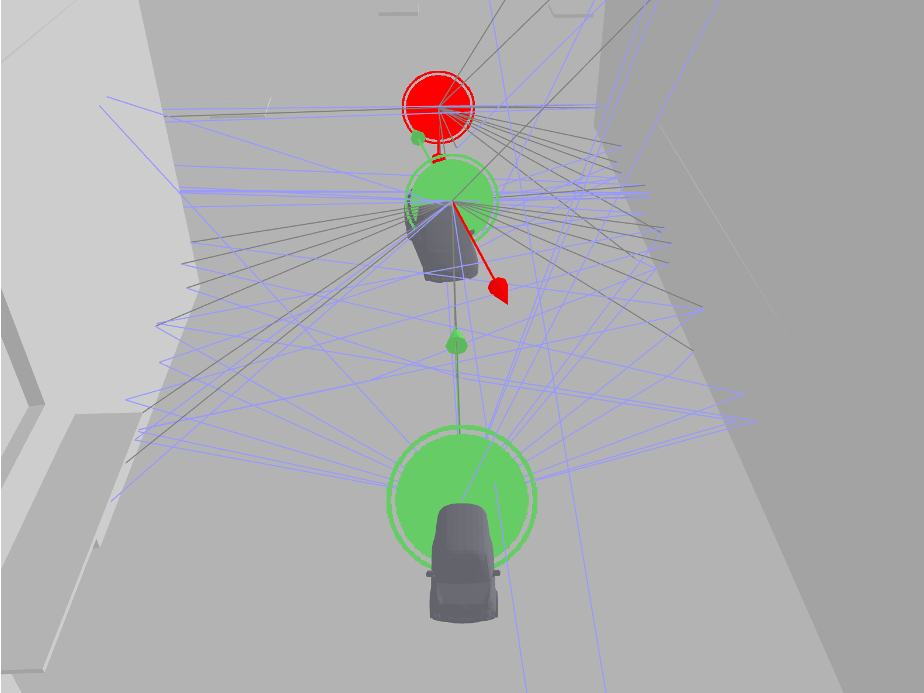} 
        \caption{Example of ray-traced propagation rays.}
        \vspace{5pt}
        \label{fig:sionna-poc2}
    \end{subfigure}
    \caption{Real picture (a), the corresponding VaN3Twin ray tracing representation (b), and an example of the computed propagation rays (c) of the road segment used in the experiments on the Ookayama campus of the Institute of Science Tokyo, Japan.}
    \label{fig:scenario-introduction}
\end{figure*}

\begin{table}
\centering
\caption{VaN3Twin settings for ray tracing.}
\label{table:validazione-parameters}
\begin{tabular}{r c}
\noalign{\smallskip}
\textbf{RT Settings} & \textbf{Value} \\
\noalign{\smallskip}
\hline
\noalign{\smallskip \smallskip} %\smallskip
    Carrier Frequency            & 60~GHz \\
    %\smallskip
    Buildings and Roads Material & Concrete~\cite{iturp2040} \\
    %\smallskip
    Trees Material               & Wood~\cite{iturp2040} \\
    %\smallskip
    Cars Material                & Metal~\cite{iturp2040} \\
    %\smallskip
    Type of Antennas            & Panasonic WiGig RSU \\
    %\smallskip
    Polarization                & Vertical \\
    %\smallskip
    Digital Model DI            & DI~5 (see Tab.~\ref{tab:di}) \\
\noalign{\smallskip}
\hline
\end{tabular}
\end{table}

The evaluation scenario involves three connected vehicles (Figure~\ref{fig:real-poc}) that interact with the Mobility DT through a $2.4$~GHz Wi-Fi \gls{cp}, and are each equipped with a roof-mounted Panasonic WiGig \gls{rsu} forming the $60$~GHz high-capacity \gls{up}. The Mobility DT and VaN3Twin are deployed at the edge---corresponding to the location of Vehicle~3---and exchange data through a wired Ethernet \gls{lan}. Figure~\ref{fig:sionna-poc} illustrates the corresponding VaN3Twin ray tracing scenario of the considered street in Ookayama, Tokyo, Japan. 
Buildings, vehicles, and antennas are modeled in VaN3Twin using material properties and radiation patterns consistent with the experimental setup (see Table~\ref{table:validazione-parameters}).
VaN3Twin ray tracing details are reported in Table~\ref{table:validazione-parameters}. %and an example of the computed propagation rays is shown in Figure~\ref{fig:sionna-poc2}. 
During the experimental campaign, the vehicles perform \texttt{iperf}-based data transmissions over the WiGig \gls{up}.
Vehicle~1 (a Toyota COMS) and Vehicle~3 (a motorized wheel cart) remain stationary and maintain a direct \gls{los} link, while Vehicle~2 (a Toyota Estima) follows a predefined trajectory (Figure~\ref{fig:sionna-poc}) that intermittently blocks the \gls{los} path.
The Mobility DT is used to predict, in real time, the $60$~GHz channel between Vehicles~1 and~3 via VaN3Twin, based on the predicted trajectory of Vehicle~2 with prediction horizon $h=500$~ms. Such predictions may be employed, for example, for proactive network topology management.
The Mobility DT simultaneously collects from each vehicle its current position and the field-measured \glspl{rssi}, which are employed for in-the-loop calibration of the prediction process. After computation, a control message is sent to the vehicles through the Wi-Fi \gls{cp}.
To quantify the effectiveness of this real-time predictive loop, we evaluate the accuracy of the \gls{rssi} forecasts produced by VaN3Twin along the trajectory of Vehicle~2. The Mobility DT employs the simple predictor for short-term trajectory estimation presented in Section~\ref{sec:mobility-channel-prediction}, while VaN3Twin operates using the DI~5 configuration reported in Table~\ref{tab:di}.
Results show a maximum average \gls{rssi} prediction error of $1.01$~dB and a 95th-percentile error of $3.05$~dB, demonstrating accurate prediction of the $60$~GHz link dynamics under mobility-induced obstruction. 
Since VaN3Twin operates on predicted positions, channel accuracy is jointly limited by trajectory prediction error and by the selected ray tracing \gls{di}, which directly determines the end-to-end latency and thus the minimum prediction horizon required for real-time operation.

\subsection{End-to-End System Latency}
Building on the formal definition introduced in Section~\ref{sec:complexity-and-time}, this section reports on-field measurements of the end-to-end latency $\tau_{\mathrm{e2e}}$, decomposed according to~\eqref{eq:e2e-latency}. %All the measured fixed latency components are summarized in Table~\ref{tab:fixed-latency}, while the resulting end-to-end latency distributions as a function of the DI level are shown in Figure~\ref{fig:e2e-latency}.

\textbf{Fixed system latency components}: are summarized in Table~\ref{tab:fixed-latency} and constitute a fixed system latency budget, independent of the DI configuration.
Measurement acquisition ($\tau_m$) was evaluated by collecting location and channel samples over $60$~s with a $100$~ms sampling interval, yielding approximately $600$ samples per experiment. No statistically significant differences were observed between static and dynamic conditions. Communication-related delays ($\tau_w$ and $\tau_{req}$) were measured using ICMP-based \gls{rtt} probing over $100$ messages and conservatively converted to one-way latency by halving the resulting value. Mobility DT update time and trajectory prediction latency ($\tau_{tp}$) was obtained by profiling the time from the scenario update up to the execution of the linear prediction algorithm described in Section~\ref{sec:mobility-channel-prediction}.

\textbf{Ray tracing execution times ($\tau_{rt}$)}:
were measured for the five DI configurations defined in Table~\ref{tab:di} using VaN3Twin with Sionna RT v1.2.1. Experiments were conducted on a workstation with an NVIDIA GeForce RTX~3090 GPU and Intel Core i9-10980XE CPUs running Ubuntu~22.04.5. $\tau_{rt}$ increases monotonically with the DI level, reflecting the higher computational load induced by denser ray interactions and more complex propagation mechanisms.

\begin{figure}
    \centering
    \includegraphics[width=1\linewidth]{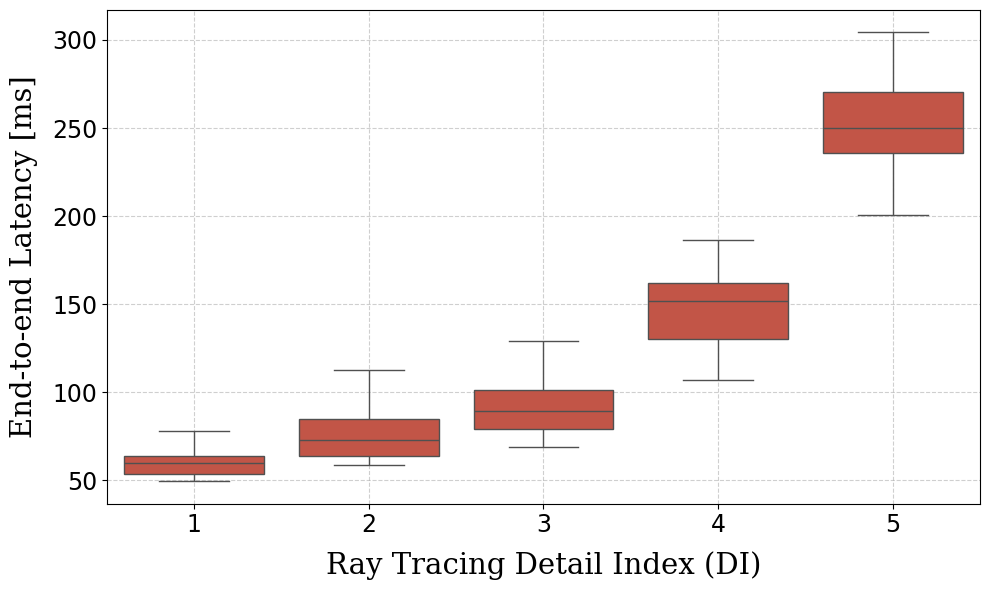}
    \caption{Distribution of the measured end-to-end latency $\tau_{e2e}$ for different Ray Tracing DI configurations.}
    \label{fig:e2e-latency}
\end{figure} 

\begin{table}
\centering
\caption{Fixed latency components contributing to $\tau_{e2e}$.}
\begin{tabular}{l c}
\noalign{\smallskip}
\textbf{Component} & \textbf{Resulting average} \\
\noalign{\smallskip}
\hline
\noalign{\smallskip } %\smallskip
    Measurement acquisition $\tau_m$      & $8.9$~ms \\
    %\smallskip
    Wi-Fi CP communication $\tau_w$             & $1.1$~ms \\
    %\smallskip
    Trajectory prediction $\tau_{tp}$     & $4.4$~ms \\
    \smallskip
    DT--VaN3Twin communication $\tau_{req}$     & $0.6$~ms\\
\hline
\noalign{\smallskip}
\textbf{Total fixed latency} & $\mathbf{16.2}$~\textbf{ms} \\
\noalign{\smallskip}
\hline
\end{tabular}
\label{tab:fixed-latency}
\end{table}

\textbf{End-to-end latency $\tau_{\mathrm{e2e}}$}: is obtained by combining the fixed system latency components with the DI-dependent ray tracing execution time. The distributions of $\tau_{\mathrm{e2e}}$ in Figure~\ref{fig:e2e-latency} show increasing latency and variability with ray tracing complexity, peaking at an average $250$~ms for DI~5. 

These results demonstrate that improving channel modeling fidelity by increasing the \gls{di} level inevitably enlarges the end-to-end latency budget. Since channel prediction must be available before transmission, this latency directly translates into a minimum required trajectory prediction horizon. As a result, higher-fidelity channel modeling forces the \gls{dt} to rely on longer-term mobility forecasts.

\subsection{Impact of trajectory prediction error}
Because the end-to-end latency of the \gls{dt} determines how far into the future vehicle positions must be predicted, longer latencies inevitably exacerbate trajectory prediction errors. This section quantifies how such latency-induced positional inaccuracies propagate into ray-traced channel prediction.

Let $\mathbf{p}(t) = [x(t),\,y(t)]^{\top}$ denote the ground-truth position at time $t$, and $\tilde{\mathbf{p}}(t) = \mathbf{p}(t) + \mathbf{e}(t)$ its perturbed counterpart, where $\mathbf{e}(t)$ models the prediction error. A normalized parameter $k \in [0,1]$ controls the maximum error magnitude through $\varepsilon_k = k\, \varepsilon_{\max}$, with $\varepsilon_{\max}=1$~m in the considered setup.  
The displacement vector is sampled uniformly from a circular annulus
\(
    \mathbf{e}(t) \sim
    \mathcal{A}\!\left( \tfrac{1}{3}\varepsilon_k,\; \tfrac{1}{2}\varepsilon_k \right),
\)
and $\mathbf{e}(t)$ is isotropically sampled within the annulus itself.
To quantify the impact of trajectory prediction inaccuracies, we evaluate two complementary metrics: (i) the deviation in received signal strength and (ii) the stability of the geometric visibility state. Both metrics are computed by comparing the channels obtained from the perturbed positions $\tilde{\mathbf{p}}(t)$ with those obtained from the true positions $\mathbf{p}(t)$.

\begin{figure} [b]
    \centering
    \includegraphics[width=\linewidth]{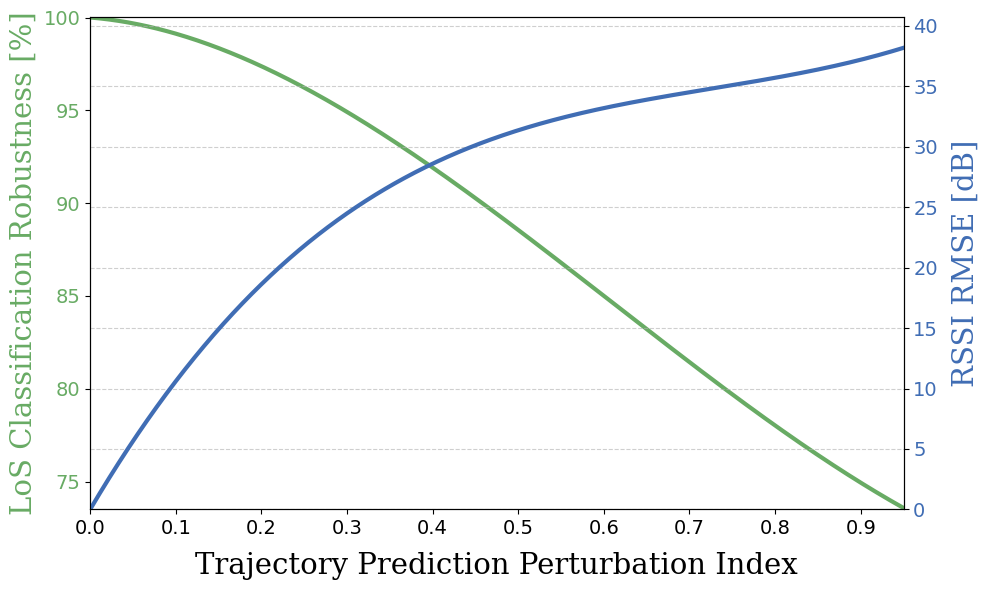}
    \caption{Experimental RSSI RMSE (blue) and LoS Prediction Accuracy (green) trends for different trajectory prediction perturbation.}
    \label{fig:trajectory-error-analysis}
\end{figure}

\textbf{RSSI Prediction Error}: let $\mathrm{RSSI}(t)$ denote the \gls{rssi} computed from $\mathbf{p}(t)$, and let $\widetilde{\mathrm{RSSI}}(t)$ denote the value obtained from $\tilde{\mathbf{p}}(t)$.  
The discrepancy induced by a perturbation index $k$ is measured through the \gls{rmse}:
\begin{align}
    \mathrm{RMSE}_k = 
    \sqrt{
    \frac{1}{N}
    \sum_{t=1}^{N}
        \left(
            \widetilde{\mathrm{RSSI}}(t) - \mathrm{RSSI}(t)
        \right)^2 },
\end{align}
where $N$ is the number of channel realizations.
Figure~\ref{fig:trajectory-error-analysis} shows the dependence of the computed \gls{rmse} (blue) on the perturbation level $k$, focusing on measurements collected during the LoS to non-LoS transition caused by the obstructing vehicle (Vehicle~2). The relationship is clearly non-linear, with a rapid increase in error for small positional deviations. For the considered scenario, $k \approx 0.1$ already yields an \gls{rmse} of approximately $20$~dB in the \gls{rssi} prediction. 
Since small geometric inaccuracies already induce large prediction errors, these results also foreshadow the sensitivity of the \gls{los} classification, analyzed next.

\textbf{LoS Classification Robustness}:
trajectory perturbations may also change the geometric visibility between a transmitter and a receiver. 
Let $\mathrm{LoS}_{\mathrm{ref}}(t)$ denote the visibility classification obtained from $\mathbf{p}(t)$ and $\widetilde{\mathrm{LoS}}(t)$ that obtained from $\tilde{\mathbf{p}}(t)$.  
The robustness of the visibility classification under perturbation level $k$ is expressed by
\begin{align}
    \eta_k = \frac{\mathrm{TP}_k + \mathrm{TN}_k}{N},
\end{align}
where $\mathrm{TP}_k$ counts the instants for which both classifications report a LoS condition and $\mathrm{TN}_k$ counts the instants for which both report a non-LoS condition. The metric $\eta_k$ therefore quantifies the agreement between the reference and perturbed LoS classifications under a positional perturbation of magnitude $k$.
Figure~\ref{fig:trajectory-error-analysis} shows $\eta_k$ (green) as a function of $k$. The geometric visibility prediction accuracy decreases monotonically, indicating a steady degradation of classification reliability with increasing positional error. For $k<0.1$, the accuracy remains close to $100\%$, whereas it degrades substantially as $k$ increases. At the maximum considered error, $\varepsilon_k \approx 1$~m ($k=1$), the consistency drops to about $75\%$. 
%In conclusion, positional errors modify the geometric blockage induced by Vehicle~2, leading to large deviations in both \gls{rssi} and LoS classification; consequently, sub-meter trajectory accuracy is required for reliable prediction, highlighting the trade-off between trajectory accuracy and the longer prediction horizons needed to accommodate the \gls{dt} latency.

In conclusion, positional errors induced by longer prediction horizons offset the gains of high-fidelity ray tracing, exposing a fundamental trade-off between channel modeling precision and end-to-end latency.

\section{Conclusions} \label{sec:conclusion}
In this work we presented and experimentally validated a real-time, predictive \gls{ndt} for V2X communications, explicitly designed around end-to-end latency constraints. By integrating a live Mobility Digital Twin with the VaN3Twin engine, we implemented a closed-loop workflow capable of operating in synchrony with the physical environment.
A key outcome of this study is the end-to-end latency analysis of the Digital Twin pipeline, covering sensing, mobility prediction, ray-tracing-based channel computation, and data exchange. This analysis allowed us to identify how the computational complexity of the physical model directly constrains the feasible prediction horizon and, consequently, the usefulness of the Digital Twin for real-time operation.
Experimental results obtained in a live urban environment demonstrate that high-fidelity channel modeling can be executed within a bounded time budget, provided that model complexity and scheduling are jointly designed. Overall, this work shows that real-time integration of mobility Digital Twins and ray-tracing-based network modeling is feasible and practically deployable.
Future work will extend this latency-aware framework toward real-time prediction of higher-layer network performance metrics leveraging VaN3Twin full-stack V2X simulation capability.

\section*{Acknowledgment}
This work was partially supported by the European Union - Next Generation EU under the Italian National Recovery and Resilience Plan (NRRP), Mission 4, Component 2, Investment 1.3, CUP D43C22003080001, partnership on “Telecommunications of the Future” (PE00000001 - program “RESTART”). This work was also supported by Japan Science and Technology Agency (JST) as part of Adopting Sustainable Partnerships for Innovative Research Ecosystem (ASPIRE), Grant Number JPMJAP2517.

\bibliographystyle{IEEEtran}
\bibliography{biblio.bib}

@article{sionna,
 title = {{Sionna: An Open-Source Library for Next-Generation Physical Layer Research}},
 author = {Hoydis, Jakob and Cammerer, Sebastian and {Ait Aoudia}, Fayçal and Vem, Avinash and Binder, Nikolaus and Marcus, Guillermo and Keller, Alexander},
 year = {2022},
 month = {Mar.},
 journal = {arXiv preprint},
 online = {https://arxiv.org/abs/2203.11854}
}

@article{ms-van3t-journal-2024,
	title = {{ms-van3t: An integrated multi-stack framework for virtual validation of V2X communication and services}},
	journal = {Computer Communications},
	volume = {217},
	pages = {70-86},
	year = {2024},
	issn = {0140-3664},
	doi = {https://doi.org/10.1016/j.comcom.2024.01.022},
	author = {F. Raviglione and C.M. Risma Carletti and M. Malinverno and C. Casetti and C.F. Chiasserini},
}

@misc{6g,
      title={Towards mmWave {V2X in 5G }and Beyond to Support Automated Driving}, 
      author={Kei Sakaguchi and Ryuichi Fukatsu and Tao Yu and Eisuke Fukuda and Kim Mahler and Robert Heath and Takeo Fujii and Kazuaki Takahashi and Alexey Khoryaev and Satoshi Nagata and Takayuki Shimizu},
      year={2020},
      eprint={2011.09590},
      archivePrefix={arXiv},
      primaryClass={cs.SI}
}

@misc{osm,
   author = {{OpenStreetMap contributors}},
   title = {{Planet dump retrieved from https://planet.osm.org }},
   howpublished = "\url{ https://www.openstreetmap.org}",
   year = {2017},
 }

@misc{iturp2040,	
    author = {{ITU-R P.2040-1}},
	year = {2015},
	month = {Jul.},
	title = {{Effects of building materials and structures on radiowave propagation above about 100 MHz}},
}

@INPROCEEDINGS{CazzellaV2XDT,
  author={Cazzella, Lorenzo and Linsalata, Francesco and Magarini, Maurizio and Matteucci, Matteo and Spagnolini, Umberto},
  booktitle={2024 IEEE 100th Vehicular Technology Conference (VTC2024-Fall)}, 
  title={A Multi-Modal Simulation Framework to Enable Digital Twin-based V2X Communications in Dynamic Environments}, 
  year={2024},
  volume={},
  number={},
  pages={1-6},
  doi={10.1109/VTC2024-Fall63153.2024.10757947}}

@misc{roongpraiwan2025digital,
title={Digital Twin-Enabled Blockage-Aware Dynamic mmWave Multi-Hop V2X Communication}, 
      author={Supat Roongpraiwan and Zongdian Li and Tao Yu and Kei Sakaguchi},
      year={2025},
      eprint={2503.03590},
      archivePrefix={arXiv},
      primaryClass={cs.NI},
      url={https://arxiv.org/abs/2503.03590}, 
}

@INPROCEEDINGS{Pegu2505:Toward,
AUTHOR="Roberto Pegurri and Francesco Linsalata and Eugenio Moro and Jakob Hoydis
and Umberto Spagnolini",
TITLE="Toward Digital Network Twins: Integrating Sionna {RT} in ns-3 for {6G}
{Multi-RAT} Networks Simulations",
BOOKTITLE="IEEE INFOCOM WKSHPS: Digital Twins over NextG Wireless Networks (DTWIN
2025) (INFOCOM DTWIN 2025)",
ADDRESS="London, United Kingdom (Great Britain)",
PAGES="5.88",
DAYS=17,
MONTH=may,
YEAR=2025
}

@ARTICLE{Noma,  author={Budhiraja, Ishan and Kumar, Neeraj and Tyagi, Sudhanshu and Tanwar, Sudeep and Han, Zhu and Piran, Md. Jalil and Suh, Doug Young},  journal={IEEE Access},   title={A Systematic Review on {NOMA} Variants for {5G} and Beyond},   year={2021},  volume={9},  number={},  pages={85573-85644},  doi={10.1109/ACCESS.2021.3081601}}

@misc{v2x,
  author = {{3GPP 38.885 V16.0.0}},
  year = {2019},
  month = {Mar.},
   title={3rd Generation Partnership Project; Technical Specification NR; Study on Vehicle-to-Everything ( Release 16)},
}

@TechReport{itu,
  author = {ITU-R},
  number={P.526},
  title={Propagation by Diffraction}, 
  institution = {International Telecommunication Union Radio Sectort(ITU-R)},
  year={2019}}

@ARTICLE{sidelink,
  author={S. {Lien} and D. {Deng} and C. {Lin} and H. {Tsai} and T. {Chen} and C. {Guo} and S. {Cheng}},
  journal={IEEE Access}, 
  title={3GPP NR Sidelink Transmissions Toward 5G V2X}, 
  year={2020},
  volume={8},
  number={},
  pages={35368-35382},
  doi={10.1109/ACCESS.2020.2973706}}

@ARTICLE{dsrc,
  author={K. {Abboud} and H. A. {Omar} and W. {Zhuang}},
  journal={IEEE Transactions on Vehicular Technology}, 
  title={Interworking of DSRC and Cellular Network Technologies for V2X Communications: A Survey}, 
  year={2016},
  volume={65},
  number={12},
  pages={9457-9470},
  doi={10.1109/TVT.2016.2591558}}

@article{RUZNIETO2023100964,
title = {{A 3D simulation framework with ray-tracing propagation for LoRaWAN communication}},
journal = {Internet of Things},
volume = {24},
pages = {100964},
year = {2023},
issn = {2542-6605},
doi = {https://doi.org/10.1016/j.iot.2023.100964},
author = {Andres Ruz-Nieto and Esteban Egea-Lopez and Jose-Marıa Molina-Garcıa-Pardo and Jose Santa},
}

@INPROCEEDINGS{matlab-rt-colosseum,
  author={Rusca, Riccardo and Raviglione, Francesco and Casetti, Claudio and Giaccone, Paolo and Restuccia, Francesco},
  booktitle={2023 Joint European Conference on Networks and Communications \& 6G Summit (EuCNC/6G Summit)}, 
  title={{Mobile RF Scenario Design for Massive-Scale Wireless Channel Emulators}}, 
  year={2023},
  volume={},
  number={},
  pages={675-680},
  doi={10.1109/EuCNC/6GSummit58263.2023.10188319}}

@techreport{ns3-sionna-falko,
    author = {Zubow, Anatolij and Pilz, Yannik and R{\"{o}}sler, Sascha and Dressler, Falko},
    doi = {10.48550/arXiv.2412.20524},
    title = {{Ns3 meets Sionna: Using Realistic Channels in Network Simulation}},
    institution = {arXiv},
    month = {12},
    number = {2412.20524},
    type = {cs.NI},
    year = {2024},
   }

@incollection{gaugel2012accurate,
  title={Accurate simulation of wireless vehicular networks based on ray tracing and physical layer simulation},
  author={Gaugel, Tristan and Reichardt, Lars and Mittag, Jens and Zwick, Thomas and Hartenstein, Hannes},
  booktitle={High Performance Computing in Science and Engineering'11: Transactions of the High Performance Computing Center, Stuttgart (HLRS) 2011},
  pages={619--630},
  year={2012},
  publisher={Springer}
}

@article{zhu2024toward,
  title={Toward Real-Time Digital Twins of EM Environments: Computational Benchmark for Ray Launching Software},
  author={Zhu, Michele and Cazzella, Lorenzo and Linsalata, Francesco and Magarini, Maurizio and Matteucci, Matteo and Spagnolini, Umberto},
  journal={IEEE Open Journal of the Communications Society},
  year={2024},
  publisher={IEEE}
}

@INPROCEEDINGS{Twardokus2505:DT-CoVeSS,
AUTHOR="Geoff Twardokus and Hanif Rahbari",
TITLE="{DT-CoVeSS: Advancing NextG C-V2X Security Evaluation through High-Fidelity Digital Twinning}",
BOOKTITLE="IEEE INFOCOM WKSHPS: Digital Twins over NextG Wireless Networks (DTWIN 2025) (INFOCOM DTWIN 2025)",
ADDRESS="London, United Kingdom (Great Britain)",
DAYS=19,
MONTH=may,
YEAR=2025
}

@ARTICLE{RAVEN-paper,
  author={Elloumi, Mohamed and Kaddoum, Georges and Zoheb Hassan, Md. and Selim, Bassant},
  journal={IEEE Internet of Things Journal}, 
  title={Digital-Twin-Empowered Interference Management for Multihop Internet of Vehicles Networks Over Millimeter Wave Bands}, 
  year={2025},
  volume={12},
  number={11},
  pages={17807-17827},
  doi={10.1109/JIOT.2025.3540750}}

@ARTICLE{TuST-paper-journal,
  author={Rapelli, Marco and Casetti, Claudio and Gagliardi, Giandomenico},
  journal={IEEE Transactions on Mobile Computing}, 
  title={Vehicular Traffic Simulation in the City of Turin From Raw Data}, 
  year={2022},
  volume={21},
  number={12},
  pages={4656-4666},
  doi={10.1109/TMC.2021.3075985}}

@INPROCEEDINGS{V2C-ADAS,
  author={Wang, Ziran and Liao, Xishun and Zhao, Xuanpeng and Han, Kyungtae and Tiwari, Prashant and Barth, Matthew J. and Wu, Guoyuan},
  booktitle={2020 IEEE 91st Vehicular Technology Conference (VTC2020-Spring)}, 
  title={A Digital Twin Paradigm: Vehicle-to-Cloud Based Advanced Driver Assistance Systems}, 
  year={2020},
  volume={},
  number={},
  pages={1-6},
  doi={10.1109/VTC2020-Spring48590.2020.9128938}}

@INPROCEEDINGS{9861008,
  author={Rapelli, Marco and Pannu, Gurjashan Singh and Dressler, Falko and Casetti, Claudio},
  booktitle={2022 IEEE 95th Vehicular Technology Conference: (VTC2022-Spring)}, 
  title={Content Sharing in Pedestrian-based Micro Clouds}, 
  year={2022},
  volume={},
  number={},
  pages={1-6},
  doi={10.1109/VTC2022-Spring54318.2022.9861008}}

@ARTICLE{10644093,
  author={Italiano, Lorenzo and Camajori Tedeschini, Bernardo and Brambilla, Mattia and Huang, Huiping and Nicoli, Monica and Wymeersch, Henk},
  journal={IEEE Communications Surveys \& Tutorials}, 
  title={A Tutorial on 5G Positioning}, 
  year={2025},
  volume={27},
  number={3},
  pages={1488-1535},
  doi={10.1109/COMST.2024.3449031}}

@ARTICLE{9839640,
  author={Hui, Linbo and Wang, Mowei and Zhang, Liang and Lu, Lu and Cui, Yong},
  journal={IEEE Network}, 
  title={Digital Twin for Networking: A Data-Driven Performance Modeling Perspective}, 
  year={2023},
  volume={37},
  number={3},
  pages={202-209},
  doi={10.1109/MNET.119.2200080}}

@ARTICLE{9429703,
  author={Wu, Yiwen and Zhang, Ke and Zhang, Yan},
  journal={IEEE Internet of Things Journal}, 
  title={Digital Twin Networks: A Survey}, 
  year={2021},
  volume={8},
  number={18},
  pages={13789-13804},
  doi={10.1109/JIOT.2021.3079510}}

@ARTICLE{10443037,
  author={Wang, Kui and Li, Zongdian and Nonomura, Kazuma and Yu, Tao and Sakaguchi, Kei and Hashash, Omar and Saad, Walid},
  journal={IEEE Transactions on Intelligent Vehicles}, 
  title={Smart Mobility Digital Twin Based Automated Vehicle Navigation System: A Proof of Concept}, 
  year={2024},
  volume={9},
  number={3},
  pages={4348-4361},
  doi={10.1109/TIV.2024.3368109}}

@misc{pegurri2025van3twinmultitechnologyv2xdigital,
      title={VaN3Twin: the Multi-Technology V2X Digital Twin with Ray-Tracing in the Loop}, 
      author={Roberto Pegurri and Diego Gasco and Francesco Linsalata and Marco Rapelli and Eugenio Moro and Francesco Raviglione and Claudio Casetti},
      year={2025},
      eprint={2505.14184},
      archivePrefix={arXiv},
      primaryClass={cs.NI},
      url={https://arxiv.org/abs/2505.14184}, 
}

@INPROCEEDINGS{11038674,
  author={Higuchi, Takamasa and Takanashi, Masaki and Sasaki, Kengo and Taguchi, Yuma and Sanda, Katsushi},
  booktitle={2025 IEEE International Conference on Pervasive Computing and Communications Workshops and other Affiliated Events (PerCom Workshops)}, 
  title={Can Vehicles Foresee Communication Disruptions?: Feasibility of V2X Network Digital Twin}, 
  year={2025},
  volume={},
  number={},
  pages={633-636},
  doi={10.1109/PerComWorkshops65533.2025.00150}}

\vfill
\end{document}